\documentclass[journal,draftcls,onecolumn,12pt,twoside]{IEEEtran}

\normalsize

\ifCLASSINFOpdf

\else

\fi
\usepackage{url}
\usepackage{graphicx}
\usepackage{amsmath,amssymb,amscd}
\usepackage{graphicx,cite}
\usepackage{algorithm}
\usepackage{algpseudocode}
\usepackage{subfigure}
\usepackage{multirow}
\usepackage{amsmath,systeme}
\usepackage{mathtools}
\usepackage{xcolor}

\begin{document}
%
\title{Cell-Free MIMO Perceptive Mobile Networks: Cloud vs. Edge Processing}

\author{Seongah Jeong, Jinkyu Kang, Osvaldo Simeone, and Shlomo Shamai (Shitz)
\thanks{The work of S. Jeong was supported by the National Research Foundation of Korea (NRF) grant funded by the Korea government (MSIT) (No. 2023R1A2C2005507).  The work of O. Simeone was  supported by the European Union’s Horizon Europe project CENTRIC (101096379), by the Open Fellowships of the EPSRC (EP/W024101/1),  by the EPSRC project (EP/X011852/1), and by Project REASON, a UK Government funded project under the Future Open Networks Research Challenge (FONRC) sponsored by the Department of Science Innovation and Technology (DSIT).  The work of S. Shamai (Shitz) was supported by the German Research Foundation (DFG) via the German-Israeli Project Cooperation (DIP), under Project SH 1937/1-1. (\textit{Corresponding author: Jinkyu Kang)}} 
\thanks{Seongah Jeong is with the School of Electronics Engineering, Kyungpook National University, Daegu, South Korea (e-mail: seongah@knu.ac.kr).}
\thanks{Jinkyu Kang is with the Department of Information and Communication Engineering, Myongji University, Gyeonggi-do, South Korea (e-mail: jkkang@mju.ac.kr).}
\thanks{Osvaldo Simeone is with the King's Communications, Learning, and Information Processing lab within the Centre for Intelligent Information Processing Systems, King's College London, London WC2B 4BG, U.K. (e-mail: osvaldo.simeone@kcl.ac.uk).}
\thanks{Shlomo Shamai (Shitz) is with the Department of Electrical and Computer Engineering, Technion, Haifa 3200003, Israel (e-mail: sshlomo@ee.technion.ac.il).}
}
\maketitle

\begin{abstract}   
Perceptive mobile networks implement sensing and communication by  reusing  existing cellular infrastructure. Cell-free multiple-input multiple-output, thanks to the cooperation among distributed access points, supports the deployment of  multistatic radar sensing, while providing high spectral efficiency for data communication services. To this end, the distributed access points communicate over fronthaul links with a central processing unit acting as a cloud processor. This work explores four different types of PMN uplink solutions based on Cell-free multiple-input multiple-output, in which the sensing and decoding functionalities are carried out at either cloud or edge. Accordingly, we investigate and compare joint cloud-based decoding and sensing (CDCS), hybrid cloud-based decoding and edge-based sensing (CDES), hybrid edge-based decoding and cloud-based sensing (EDCS) and edge-based decoding and sensing (EDES). In all cases, we target a unified design problem formulation whereby the fronthaul quantization of signals received in the training and data phases are jointly designed to maximize the achievable rate under sensing requirements and fronthaul capacity constraints. Via numerical results, the four implementation scenarios are compared as a function of the available fronthaul resources by highlighting the relative merits of edge- and cloud-based sensing and communications. This study provides guidelines on the optimal functional allocation in fronthaul-constrained networks implementing integrated sensing and communications.
\end{abstract}


\IEEEpeerreviewmaketitle

\section{Introduction}\label{sec:intro}
\subsection{Context and Motivation}
For 6G and beyond, perceptive mobile networks (PMNs) have been introduced as a general framework that integrates sensing capability into the cellular network deployments \cite{Zhang20Aero, Zhang21TVT, Liu18TWC, Liu18TSP, Xie22Arxiv}. In that sense, the PMNs can be regarded as a special case of the recently-proposed framework of integrated sensing and communication (ISAC) \cite{Liu23, Shamai23ISIT}, joint radar and communication (JRC) \cite{Feng20CM} or joint communication and radar/radio sensing (JCAS) \cite{Chiriyath17TCCN, Zhang19TVT}, known as dual-function radar communications to focus on integration of sensing and communication into one system. By suitably reusing hardware, spectral resources, and signals of the underlying mobile network, radar sensing services can be potentially provided with minimal degradation in performance in the communication services \cite{Ngo24Arxiv}. This approach is envisaged to support location-based services and applications in 6G networks. The introduction of PMNs is also  facilitated by the adoption of higher carrier frequencies in 6G such as in the millimeter wave (mmWave) and terahertz (THz) bands, and by the shift towards an open, software-based, network architecture \cite{Polese23ComTu}. 

In cell-free multiple-input multiple-output (CF-MIMO) architectures, multiple access points (APs) collaborate for the provision of communication services  by communicating over a fronthaul network with a  central processing unit (CPU). CF-MIMO is well aligned with emerging specifications such as the open radio access network (O-RAN) architecture \cite{Polese23ComTu}. PMNs based on CF-MIMO deployments are particularly promising, as the coherent processing across multiple APs can facilitate radar sensing tasks such as target detection \cite{JSA16ETT}. 
 
The main purpose of this work is to explore the functional split between APs and CPU  in PMN systems based on the CF-MIMO architectures. We aim at  developing an understanding about the pros and cons of possible PMN deployment scenarios in which sensing and communication functionalities are implemented at either edge or cloud as a function of the level of fronthaul connectivity between APs and CPU.

\subsection{Related Works}\label{sec:rel}

 \subsubsection{Perceptive Mobile Networks} \label{sec:arc}
PMNs can have different architectures. In a \textit{monostatic dual-function radar and communication } architecture, some APs are deployed as both transmitters and receivers for the sensing signals \cite{Liu18TWC, Xie22Arxiv}. In these systems, the selected APs need to work in full-duplex mode, and inter-AP synchronization is not necessary for sensing. Alternatively, PMNs can be deployed within a  CF-MIMO  architecture \cite{Zhang21TVT, Xie22Arxiv, JSA16ETT}. In this case, the APs can be either full-duplex or half-duplex, with half-duplex APs serving as either the transmitter or the receiver of sensing signal. Finally, PMNs can also deploy  separate sensing  terminals acting as  dedicated receivers of radar  signals that have passive sensing functionalities  \cite{Xie22TWC, Huang20TWC,  Xie22Arxiv}.

 \subsubsection{Signals for Integrated Sensing and Communications}\label{sec:singals}
In PMNs, communication signals can be  reused for sensing as well \cite{Zhang21TVT}. In particular, referring to the 3GPP specification 38.211 \cite{Lin19CSM}, there are three options available for repurposing existing signals for sensing: reference signal for channel estimation that  include  demodulation reference signals (DMRS) and sounding reference signals (SRSs);  deterministic nonchannel-estimation signal, including synchronization signals (SSs) and physical broadcast channel signals (PBCHs); and data payload signals  in the physical downlink shared channel  (PDSCH) and in the physical uplink shared channel (PUSCH). Reference signals for channel estimation can be allocated during user scheduling period for both sensing and communication. The usefulness of synchronizing and broadcast signals for sensing is limited by their short length, which constrains the sensing resolution. Data payload signals are known in the downlink, while, in uplink, they need to be estimated, e.g., using the decision-directed method to remodulate the demodulated and decoded data signals, increasing the sensing complexity. However, since data payload signals are flexible, they can be easily designed for both sensing and communication.      

\subsubsection{Performance Metrics for Sensing}\label{sec:PMs}        
  The Cram\'{e}r-rao lower bound  (CRLB) is often leveraged for radar sensing optimization \cite{Gogineni15TSP, JSA14TVT}. Information-theoretic performance metrics such as Bhattacharyya distance, Kullback Leibler divergence, J-divergence, and mutual information are also adopted as an alternative to bound the probability of sensing errors in terms of missed detection, false alarm and Bayesian risk \cite{Xu15ICT, JSA16ETT, Kailath67TCOM, Kay09Aero, Stoica13TSP}.
 

 \subsection{Contributions}\label{sec:contri}
\begin{figure}[t]
\centering
\includegraphics[width=14cm]{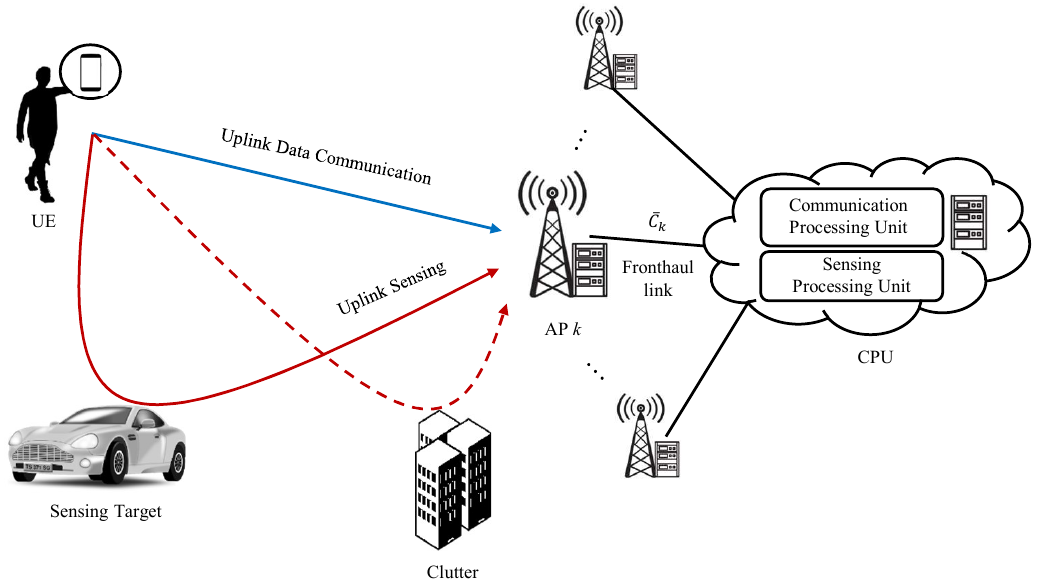}
\caption{PMNs simultaneously provide communication and uplink sensing functionalities by reusing the same uplink signals.}
\label{fig:fig1}
\end{figure}

\begin{figure}[t]
\centering
\includegraphics[width=14cm]{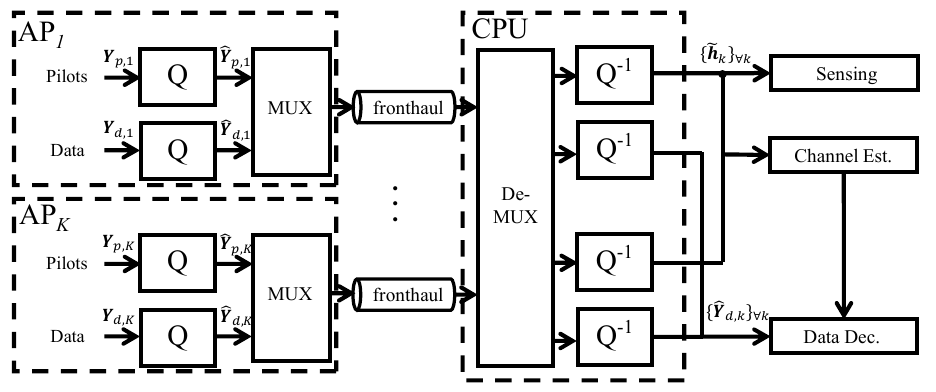}
\caption{Block diagram of joint cloud-based decoding and sensing (CDCS).}
\label{fig:fig_CDCS}
\end{figure}

\begin{figure}[t]
\centering
\includegraphics[width=14cm]{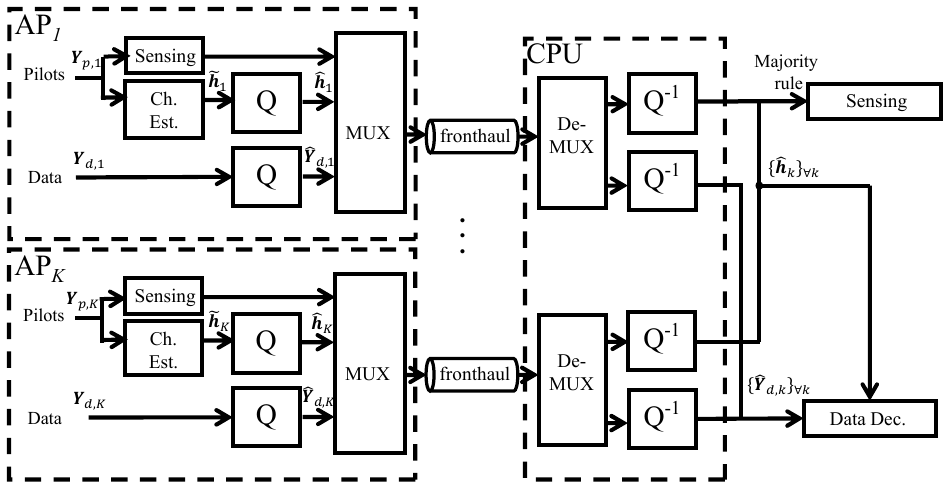}
\caption{Block diagram of hybrid cloud-based decoding and edge-based sensing (CDES).}
\label{fig:fig_CDES}
\end{figure}

\begin{figure}[t]
\centering
\includegraphics[width=14cm]{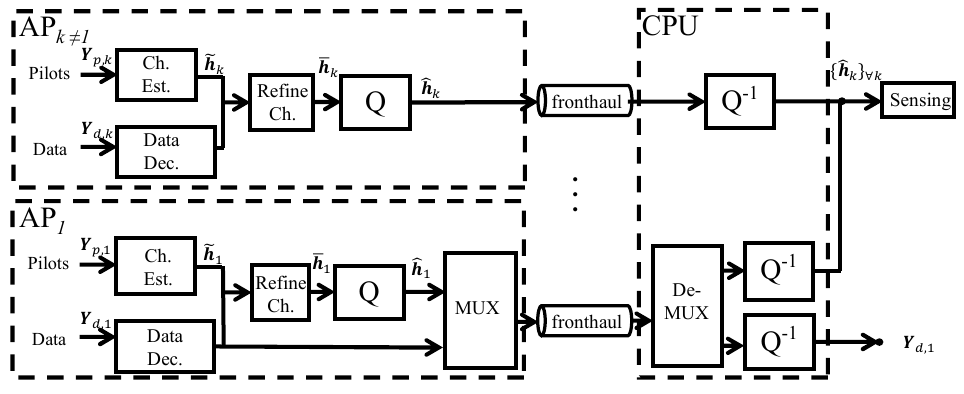}
\caption{Block diagram of hybrid edge-based decoding and cloud-based sensing (EDCS).}
\label{fig:fig_EDCS}
\end{figure}

\begin{figure}[t]
\centering
\includegraphics[width=14cm]{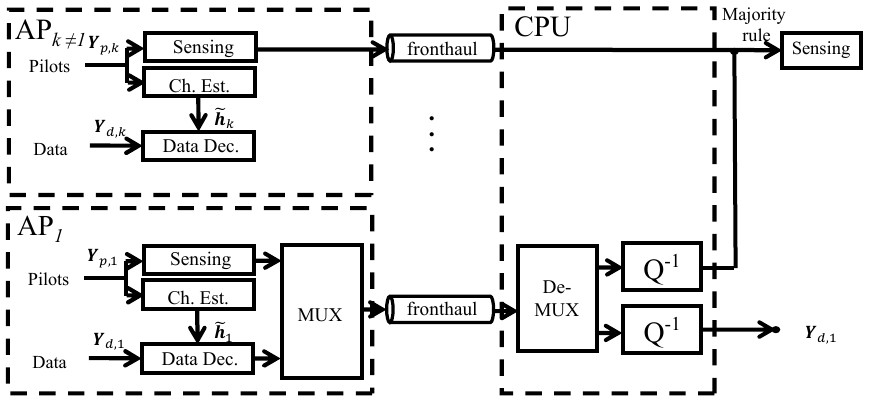}
\caption{Block diagram of edge-based decoding and sensing (EDES).}
\label{fig:fig_EDES}
\end{figure}

 As illustrated in Fig. \ref{fig:fig1}, this work studies PMNs based on a CF-MIMO architecture  consisting of multiple half-duplex APs and a CPU by focusing on the uplink. This work explores four different types of PMN uplink solutions, in which the sensing and
decoding functionalities are carried out at either cloud or edge. Specifically, we study and optimize \textit{joint cloud-based decoding and sensing} (CDCS) in Fig. \ref{fig:fig_CDCS}, \textit{hybrid cloud-based decoding and edge-based sensing} (CDES) in Fig. \ref{fig:fig_CDES}, \textit{hybrid edge-based decoding and cloud-based sensing} (EDCS) in Fig. \ref{fig:fig_EDCS} and \textit{edge-based decoding and sensing} (EDES) \ref{fig:fig_EDES}. The main contributions are as follows. 

\begin{itemize}
\item We introduce and describe CF-MIMO-based PMNs with CDCS, CDES, EDCS and EDES functional splits. Different functional splits target distinct fronthaul capacity regimes, and yield different trade-offs between the performance of  communication and sensing.

\item For all scenarios,  we target a unified design problem formulation whereby the fronthaul
quantization of signals received in the training and data phases are jointly designed to maximize the
achievable rate under sensing requirements and fronthaul capacity constraints. Specifically, the performance objectives for data communication and radar sensing are the ergodic rates  and the Bhattacharyya distance, respectively.  

\item Simulations results highlight the trade-off between data communication and radar sensing  in PMNs, and presents guidelines for the optimization of the functional splits for sending and communications. 

\end{itemize}

\begin{table}[t]
\caption{Operations carried out at the edge (APs) and at the cloud (CPU) for the four functional under study.}\label{table:abb}
\centering
\begin{tabular}{| p{1cm} | p{0.8cm} || p{6.5cm} | p{6.5cm}|} 
\hline
\multicolumn{2}{|c||}{Type} & Training phase & Data phase \\
\hline 
\hline 
{\textbf{CDCS}} & AP & $\bullet$ Quantize and forward pilot signal ${\pmb{Y}}_{p,k}$ to CPU & $\bullet$ Quantize and forward data signal ${\pmb{Y}}_{d,k}$ to CPU \\
\cline{2-4}
& CPU & $\bullet$ Receive quantized pilot signal $\hat {\pmb{Y}}_{p,k}$ & \multirow{3}{6.5cm}{$\bullet$ Decode data with all quantized data signals $\{ \hat {\pmb{Y}}_{d,k} \}_{\forall k}$ based on CPU's channel estimate $\{ \tilde {\pmb{h}}_{k} \}_{\forall k}$} \\
& & $\bullet$ Detect and decide target's presence with all quantized pilot signals $\{ \hat {\pmb{Y}}_{p,k}\}_{\forall k}$ & \\
& & $\bullet$ Estimate channels $\{ \tilde {\pmb{h}}_{k} \}_{\forall k}$ with ML & \\
\hline
{\textbf{CDES}} & AP & $\bullet$ Detect target with pilot signal ${\pmb{Y}}_{p,k}$ & $\bullet$ Quantize and forward data signal ${\pmb{Y}}_{d,k}$ to CPU \\
& & $\bullet$ Forward sensing results to CPU & \\
& & $\bullet$ Estimate channel $\tilde {\pmb{h}}_k$ with ML & \\
& & $\bullet$ Quantize and forward channel estimate $\tilde {\pmb{h}}_k$ to CPU & \\
\cline{2-4}
& CPU & $\bullet$ Decide target's presence via majority rule & \multirow{3}{6.5cm}{$\bullet$ Decode data with all quantized data signals $\{ \hat {\pmb{Y}}_{d,k} \}_{\forall k}$ based on CPU's channel estimate $\{ \hat {\pmb{h}}_{k} \}_{\forall k}$} \\
& & $\bullet$ Receive quantized channel estimate $\hat {\pmb{h}}_k$ & \\
& & & \\
\hline
{\textbf{EDCS}} & AP & $\bullet$ Estimate channel $\tilde {\pmb{h}}_k$ with ML & $\bullet$ Decode data based on AP's channel estimate $\tilde {\pmb{h}}_k$ \\
& & & $\bullet$ Refine channel estimate $\tilde {\pmb{h}}_k$ with decoded data \\
& & & $\bullet$ Quantize and forward refined channel estimate $\bar {\pmb{h}}_k$ to CPU \\
\cline{2-4}
& CPU & & $\bullet$ Receive the decoded message from AP with the largest SNR \\
& & & $\bullet$ Receive quantized refined channel estimate $\{ \hat {\pmb{h}}_k \}_{\forall k} \}$ \\
& & & $\bullet$ Detect and decide target's presence with all quantized refined AP's channel estimates $\{ \hat {\pmb{h}}_k \}_{\forall k}$ \\
\hline
{\textbf{EDES}} & AP & $\bullet$ Detect target with pilot signal ${\pmb{Y}}_{p,k}$ & $\bullet$ Decode data based on AP's channel estimate $\tilde {\pmb{h}}_k$ \\
& & $\bullet$ Forward sensing results to CPU & \\
& & $\bullet$ Estimate channel $\tilde {\pmb{h}}_k$ with ML & \\
\cline{2-4}
& CPU & $\bullet$ Decide target's presence via majority rule & $\bullet$ Receive the decoded message from AP with the largest SNR \\
\hline
\end{tabular}
\end{table}


\subsection{Organization}

The rest of the paper is organized as follows. Sec. \ref{sec:syssig} introduces the signal and channel models of PMNs, and the design goals of the four types of functional splits between communications and sensing. The architecture and detailed design elements of four schemes are provided in Sec. \ref{sec:CDCS}, Sec. \ref{sec:CDES}, Sec. \ref{sec:EDCS} and Sec. \ref{sec:EDES}. Numerical results are given in Sec. \ref{sec:num} in order to validate the trade-offs between data communication and sensing that are afforded by different functional splits. Finally, Sec. \ref{sec:con} concludes this paper.

{\textit{Notation}}: $E[\cdot]$ denotes the expectation of the argument matrix; We reserve the superscript $\pmb{A}^T$, $\pmb{A}^H$ and $\pmb{A}^{-1}$ for the transpose, the conjugate transpose and the pseudo-inverse of $\pmb{A}$, respectively; The matrix $\pmb{I}_i$ denotes the $i \times i$ identity; $H(\pmb{X})$ represents the entropy of the random variable $\pmb{X}$ to be upper-bounded as the differential entropy \cite{Cov06}; $I (\pmb{X};\pmb{Y})$ represents mutual information as the amount of uncertainty in $\pmb{X}$ due to knowledge of $\pmb{Y}$; and $I (\pmb{X};\pmb{Y}| \pmb{Z})$ represents conditional mutual information between $\pmb{X}$ and $\pmb{Y}$ conditioned with $\pmb{Z}$.   

\section{System and Signal Models}\label{sec:syssig} 
\subsection{System Model}\label{sec:sys}
As illustrated in Fig. \ref{fig:fig1}, we consider a PMN for integrated communication and radar sensing that adopts the CF-MIMO architecture. We focus on the spectral resources occupied by a single user equipment (UE), and assume that the UE's uplink signal is used to support detection of a radar target and to provide uplink data communication.

The architecture of interest consists of $K$ half-duplex APs equipped with $N_r$ antennas, while the UE is equipped with a single antenna. All APs are connected to the the CPU via orthogonal finite-capacity fronthaul links. Each $k$th fronthaul link between AP $k \in \mathcal{K}$ and CPU has capacity $\bar{C}_k$ in bits/s/Hz, which is normalized with respect to the bandwidth of the uplink channel. We note that the fronthaul links can be either wired, such as a fiber optic link, or wireless, such as a microwave link. 

\subsection{Signal Model}\label{sec:sig}
The UE transmission consists of a training slot of length $T_p$ channel uses and of a data slot of length $T_d$ channel uses, with $T=T_p+T_d$. The signal transmitted by the UE is given by a $T \times 1$ complex vector $\pmb{x}$, which is split into the $T_p \times 1$ pilot signal $\pmb{x}_p$ and the $T_d \times 1$ data signal $\pmb{x}_d$. The per-block power constraint $\|\pmb{x}\|^2/T \le P_T$, is divided as $\|\pmb{x}_p\|^2/T_p = P_p$ and $\|\pmb{x}_d\|^2/T_d = P_d$ for training and data transmission phases, respectively. The training signal is $\pmb{x}_p=\sqrt{P_p}\pmb{s}_p$, where $\pmb{s}_p$ is a $T_p \times 1$ vector of independent and identically (i.i.d) complex Gaussian $\mathcal{CN}(0,1)$ random variables. Similarly, during the data phase, the UE transmits the data streams denoted as $\pmb{x}_d=\sqrt{p_d}\pmb{s}_d$, where $\pmb{s}_d$ is a $T_d \times 1$ vector of i.i.d $\mathcal{CN}(0,1)$ variables denoting the standard random coding construction. 

\subsubsection{Channel Model}\label{sec:Ch}
We consider the presence of a single stationary target in a clutter field \cite{Zinat22Globecom, JSA16ETT}. Each AP receives a noisy version of the signal transmitted from the UE, which is reflected from the surveillance area, where a single-point target can exist. The assumption of stationary target and scatterers can be considered as the worst-case scenario for a more general set-up with Doppler effects \cite{Kay09Aero}. If the target is present, the $N_r \times 1$ channel vector from the UE to the $k$th AP is modeled as 
\begin{equation}\label{eq:ch}
\pmb{h}_k = \pmb{g}_k + \pmb{c}_k,
\end{equation}     
where $\pmb{g}_k$ and $\pmb{c}_k$ describe the target scattering effects and clutter components, respectively; while, if the target is absent, we have $\pmb{h}_k=\pmb{c}_k$. In (\ref{eq:ch}), $\pmb{g}_k$ is the random complex amplitude of the target return, which is written as \cite{Xu23ICC}
\begin{equation}\label{eq:ch_g}
\pmb{g}_k =\alpha_k\pmb{a}(\theta_k),
\end{equation}     
where $\alpha_k \sim \mathcal{CN}(0, \sigma_{\alpha,k}^2)$ represents the combined communication and sensing channel gain, which accounts for the effects of pathloss and radar cross section (RCS) of the target and follow a Swerling-I target-type model; and $\pmb{a}(\theta_k)=[1 \,\, \exp(-j\pi\sin(\theta_k)) \cdots \exp(-j\pi(N_r-1)\sin(\theta_k)]$ is the array response vector, with $\theta_k$ being the angles of arrival of the transmitting UE and receiving AP $k$ through the single target. By (\ref{eq:ch_g}), the covariance matrix of the target reflection is $\pmb{\Omega}_{g,k}=\sigma_{\alpha,k}^2\pmb{a}(\theta_k)\pmb{a}(\theta_k)^H$. The Swerling-I model can be justified by the assumptions that the fluctuations of RCS are slow and that the sensing channel does not change within the signal transmission of $\pmb{s}_p$ and $\pmb{s}_d$ \cite{Richards10Book, Rihaczek67Book}. The clutter components $\pmb{c}_k$ is modeled as having Gaussian entries distributed as $\mathcal{CN}(0, \sigma_{c,k}^2)$ by invoking the central limit theorem \cite{Kay09Aero}. 

\subsubsection{Training Phase}\label{sec:Sigp}
During the training phase, the $N_r \times T_p$ signal $\pmb{Y}_{p,k}$ received at AP $k \in \mathcal{K}$ is given as
\begin{equation}\label{eq:Sigp}
\pmb{Y}_{p,k}=\sqrt{P_p}\pmb{h}_k\pmb{s}_p^T + \pmb{Z}_{p,k},
\end{equation}
where $\pmb{Z}_{p,k}$ represents $N_r \times T_p$ matrices of i.i.d $\mathcal{CN}(0,\sigma_{z,k}^2)$ variables. Depending on the target presence or absence, we have two possible models for the received signal $\pmb{Y}_{p,k}$ at AP $k$, namely  
\begin{subequations}\label{eq:Hypo_p}
\begin{eqnarray}
\text{(AP $k$)} && \mathcal{H}_0: \pmb{Y}_{p,k}=\sqrt{P_p}\pmb{c}_k\pmb{s}_p^T +\pmb{Z}_{p,k},\\
&& \mathcal{H}_1: \pmb{Y}_{p,k}=\sqrt{P_p}\pmb{g}_k\pmb{s}_p^T + \sqrt{P_p}\pmb{c}_k\pmb{s}_p^T + \pmb{Z}_{p,k},
\end{eqnarray}
\end{subequations}
where $\mathcal{H}_0$ and $\mathcal{H}_1$ denote the hypotheses under which the target is absent and present, respectively.

\subsubsection{Data Phase}\label{sec:Sigd}
Similar with the training phase, the $N_r \times T_d$ received data signal $\pmb{Y}_{d,k}$ at AP $k$ can be written as
\begin{equation}\label{eq:Sigd}
\pmb{Y}_{d,k}=\sqrt{P_d}\pmb{h}_k\pmb{s}_d^T + \pmb{Z}_{d,k},
\end{equation}
where $\pmb{Z}_{d,k}$ represents the $N_r \times T_d$ matrices of i.i.d $\mathcal{CN}(0,\sigma_{z,k}^2)$ variables. Based on (\ref{eq:Sigd}), the received signal is modeled as 
\begin{subequations}\label{eq:Hypo_d}
\begin{eqnarray}
\text{(AP $k$)} && \mathcal{H}_0: \pmb{Y}_{d,k}=\pmb{c}_k\pmb{x}_d^T +\pmb{Z}_{d,k},\\
&& \mathcal{H}_1: \pmb{Y}_{d,k}=\pmb{g}_k\pmb{x}_d^T + \pmb{c}_k\pmb{x}_d^T + \pmb{Z}_{d,k},
\end{eqnarray}
\end{subequations}
depending on whether the target is absent, $\mathcal{H}_0$, or present, $\mathcal{H}_1$. The variables $\pmb{g}_k$, $\pmb{c}_k$, $\pmb{Z}_{p,k}$ and $\pmb{Z}_{d,k}$ for all $k \in \mathcal{K}$ are assumed to be independent for different values of $k$ and antenna elements under the assumption that the APs and antenna elements are sufficiently separated \cite{Kay09Aero}. Moreover, the second-order statistics $\pmb{\Omega}_{g,k}$, $\sigma_{c,k}^2$ and $\sigma_{z,k}^2$ are assumed to be known from prior measurements or prior information \cite{Gini08Book}.     

\subsection{Design Architectures and Goals}\label{sec:goal}
In this work, we explore four different implementations of PMNs that differ depending on the communication/sensing functional split across cloud and edge: 
\begin{itemize} 
\item \textit{Joint cloud-based decoding and sensing} (CDCS): As shown in Fig. \ref{fig:fig_CDCS}, with CDCS, both decoding and sensing are carried out at the CPU based on the compressed signals transmitted from the APs via the fronthaul. 
\item \textit{Hybrid cloud-based decoding and edge-based sensing} (CDES): As illustrated in Fig. \ref{fig:fig_CDES}, in CDES, the APs detect the presence of the sensing target based on the received signal. Then, each AP estimates the channel, and performs fronthaul compression for the estimated channel and received signal. The CPU carries out channel decoding based on the quantized estimated channels and received signals.
\item \textit{Hybrid edge-based decoding and cloud-based sensing} (EDCS): As seen in Fig. \ref{fig:fig_EDCS}, with EDCS, the APs estimate the channel, and decode the communication signal. The channel estimate is refined based on the decoded data. The channel is finally forwarded to the CPU for radar sensing.
\item \textit{Edge-based decoding and sensing} (EDES): As illustrated in Fig. \ref{fig:fig_EDES}, in EDES, both decoding and sensing are implemented at APs. 
\end{itemize}

In all cases, the primary function of PMNs is data communication, with sensing being a secondary aim. Accordingly, the aim of the system is to maximize the uplink data communication rate, while guaranteeing probability of detection requirements for the radar sensing task. 
\section{Joint Cloud-based Decoding and Sensing}\label{sec:CDCS}
In this section, we formulate and address the design problem for CDCS as illustrated in Fig. \ref{fig:fig_CDCS}. In CDCS, both decoding and radar sensing are performed at the CPU based on the signals received on the fronthaul. To this end, the CPU first performs radar sensing during the pilot phase. Then, the obtained information about the target's presence or absence is used for estimating the channel state information (CSI) based on the received pilot slot for data detection.    

\subsection{Training Phase}\label{sec:CDCSp}
On the orthogonal-access fronthaul links, each AP $k$ quantizes the received vector $\pmb{Y}_{p,k}$ in (\ref{eq:Hypo_p}), and sends a quantized version of $\pmb{Y}_{p,k}$ to the CPU. By using standard rate-distortion theory arguments \cite{GamalBook}, we model the quantization effect by means of an additive quantization noise as
\begin{equation}\label{eq:Q_CDCSp}
\hat{\pmb{Y}}_{p,k} = \pmb{Y}_{p,k} + \pmb{Q}_{p,k},
\end{equation}
where the quantization noise matrix $\pmb{Q}_{p,k} \in \mathbb{C}^{N_r \times T_p}$ is assumed to have i.i.d. $\mathcal{CN}(0,\sigma_{p,k}^2)$ entries. We suppose that the quantization noises $\pmb{Q}_{p,k}$ are independent for the different index $k$, which can be realized by using separate quantizers for the signals of the different APs \cite{GamalBook, Marce90TCOM}. Note that the Gaussian assumption of quantization noise is justified by the fact that a high-dimensional dithered lattice quantizer preceded by a linear transform can obtain a Gaussian quantization noise \cite{Gersho92, Cov06}.  

For radar sensing, the received pilot signals at CPU in (\ref{eq:ch_CDCSp}) can be written as 
\begin{subequations}\label{eq:HypoR_CDCSyp}
\begin{eqnarray}
\text{(CPU)} && \mathcal{H}_0: \hat{\pmb{Y}}_{p,k} = \sqrt{P_p}\pmb{c}_k\pmb{s}_p^T + \pmb{Z}_{p,k} +\pmb{Q}_{p,k} \\
&& \mathcal{H}_1: \hat{\pmb{Y}}_{p,k} = \sqrt{P_p}\pmb{g}_k\pmb{s}_p^T +\sqrt{P_p}\pmb{c}_k\pmb{s}_p^T + \pmb{Z}_{p,k} +\pmb{Q}_{p,k}. 
\end{eqnarray}
\end{subequations}  
For analytical convenience, the signal received at the CPU is whitened with respect to the covariance of the overall additive noise $\sqrt{P_p}\pmb{c}_k\pmb{s}_p^T + \pmb{Z}_{p,k} +\pmb{Q}_{p,k}$, leading to the equivalent model for the received signals \cite{JSA16ETT} 
\begin{subequations}\label{eq:HypoR_CDCSp}
\begin{eqnarray}
\text{(CPU)} && \mathcal{H}_0: \pmb{r}_p \sim \mathcal{CN}(\pmb{0}, \pmb{I}) \\
&& \mathcal{H}_1: \pmb{r}_p \sim \mathcal{CN}(\pmb{0}, \pmb{D}\pmb{\Lambda}\pmb{D}+\pmb{I}), 
\end{eqnarray}
\end{subequations}   
where $\pmb{r}_p = [\left(\pmb{r}_{p,1}\right)^T \cdots \left(\pmb{r}_{p,K}\right)^T]^T$ with $\pmb{r}_{p,k} = [\left(\pmb{r}_{p,k,1}\right)^T \cdots \left(\pmb{r}_{p,k,T_p}\right)^T]^T$; $\pmb{r}_{p,k,t} = \pmb{D}_{k}(\hat{\pmb{Y}}_{p,k})_t$ with $(\hat{\pmb{Y}}_{p,k})_t$ being the column vector of $\hat{\pmb{Y}}_{p,k}$, i.e., the $N_r \times 1$ signal vector at the $t$th channel use of AP $k$; $\pmb{D}_k$ is the whitening matrix for AP $k$ given by $\pmb{D}_k=\pmb{I}_{N_r}/\sqrt{P_p\sigma_{c,k}^2+\sigma_{z,k}^2+\sigma_{p,k}^2}$; $\pmb{D}=\text{diag}\{\pmb{I}_{T_p}\otimes\pmb{D}_1, \dots, \pmb{I}_{T_p}\otimes\pmb{D}_{K}\}$; and $\pmb{\Lambda} = \text{diag}\{\pmb{I}_{T_p}\otimes\pmb{\Lambda}_1, \dots, \pmb{I}_{T_p}\otimes\pmb{\Lambda}_K\}$ with $\pmb{\Lambda}_k =P_p\pmb{\Omega}_{g,k}$  is the block diagonal matrix. 

The detection problem in (\ref{eq:HypoR_CDCSp}) has the standard Neyman-Pearson solution obtained by the test 
\begin{align}\label{eq:test_CDCSp}
        							&\mathcal{H}_1 \nonumber\\[-15pt]
\text{(CPU)} \hspace{0.2cm} \pmb{r}_p^H\pmb{T}\pmb{r}_p        &\gtreqless   \nu_p, \\[-15pt]
        							&\mathcal{H}_0 \nonumber
\end{align} 
where we have defined $\pmb{T}=\pmb{D}\pmb{\Lambda}\pmb{D}(\pmb{D}\pmb{\Lambda}\pmb{D}+\pmb{I}_{KN_rT_p})^{-1}$, and $\nu_p$ is the threshold based on the tolerated false alarm probability \cite{Kay93Book}. 

For channel estimation at the CPU, the CPU adopts the maximum likelihood (ML) method, which does not require knowledge of the statistics of the channel \cite{Morelli01TSP, Nunes99Globecom}. With the received quantized signal in (\ref{eq:Q_CDCSp}), the ML estimate of $\pmb{h}_k$ is given as  
\begin{equation}\label{eq:ch_CDCSp}
\text{(CPU)} \hspace{0.2cm} \tilde{\pmb{h}}_k = \hat{\pmb{Y}}_{p,k}\frac{\pmb{x}_p^*}{\|\pmb{x}_p\|^2}= \pmb{h}_k + (\pmb{Z}_{p,k}+\pmb{Q}_{p,k})\frac{\pmb{x}_p^*}{\|\pmb{x}_p\|^2} = \pmb{h}_k + \pmb{e}_k, 
\end{equation}
for all APs $k$, where the estimation error is $\pmb{e}_{k} \in \mathbb{C}^{N_r \times 1}=(\pmb{Z}_{p,k}+\pmb{Q}_{p,k})\frac{\pmb{x}_p^*}{\|\pmb{x}_p\|^2}$, which has i.i.d. entries $\mathcal{CN}(0, (\sigma_{z,k}^2+\sigma_{p,k}^2)/(P_pT_p))$. The variance of the channel estimate $\tilde{\pmb{h}}_k$ is denoted as $\pmb{\Omega}_{\tilde{h},k|\mathcal{H}_i}(\sigma_{p,k}^2)$, and it depends on the received signal $\hat{\pmb{Y}}_{p,k}$, and therefore on the hypothesis $\mathcal{H}_i$ regarding the target presence, for $i \in \{0,1\}$. The variances of the estimated channel and the estimation error under the hypothesis $\mathcal{H}_0$ or $\mathcal{H}_1$ are obtained as    
\begin{subequations}\label{eq:ML_var_CDCS}
\begin{eqnarray}
&& \hspace{-0.8cm}\mathcal{H}_0: \pmb{\Omega}_{\tilde{h},k|\mathcal{H}_0}(\sigma_{p,k}^2) = \left(\frac{\sigma_{c,k}^2P_pT_p-\left(\sigma_{z,k}^2+\sigma_{p,k}^2\right)}{P_pT_p}\right)\pmb{I},\\
&& \hspace{-0.8cm} \mathcal{H}_1: \pmb{\Omega}_{\tilde{h},k|\mathcal{H}_1}(\sigma_{p,k}^2) = \pmb{\Omega}_{g,k}+\left(\frac{\sigma_{c,k}^2P_pT_p-\left(\sigma_{z,k}^2+\sigma_{p,k}^2\right)}{P_pT_p}\right)\pmb{I},
\end{eqnarray}
\end{subequations}  
respectively, where $\pmb{\Omega}_{h,k}=\sigma_{c,k}^2\pmb{I}_{N_r}$ for $\mathcal{H}_0$, while $\pmb{\Omega}_{h,k}=\pmb{\Omega}_{g,k}+\sigma_{c,k}^2\pmb{I}_{N_r}$ for $\mathcal{H}_1$.    

\subsection{Data Phase}\label{sec:CDCSd}
From (\ref{eq:Hypo_d}) and following the standard rate-distortion approach as in (\ref{eq:Q_CDCSp}) of training phase, the compressed data signal received at the CPU can be expressed as the sum of the useful term $\tilde{\pmb{h}}_k\pmb{x}_d^T$ and of the equivalent noise $\pmb{N}_{d,k}=\pmb{e}_k\pmb{x}_d^T+\pmb{Z}_{d,k}+\pmb{Q}_{d,k}$, i.e., 
\begin{equation}\label{eq:Q_CDCSd}
\hat{\pmb{Y}}_d = \tilde{\pmb{h}}\pmb{x}_d^T + \pmb{N}_d,
\end{equation}
where we have defined $\hat{\pmb{Y}}_{d} = [\hat{\pmb{Y}}_{d,1}^T, \hat{\pmb{Y}}_{d,2}^T, \dots, \hat{\pmb{Y}}_{d,K}^T]^T$ with $\hat{\pmb{Y}}_{d,k} = \tilde{\pmb{h}}_k\pmb{x}_d^T + \pmb{N}_{d,k}$, $\tilde{\pmb{h}} = [\tilde{\pmb{h}}_1^T, \tilde{\pmb{h}}_2^T, \dots, \tilde{\pmb{h}}_K^T]^T$ and $\pmb{N}_{d} = [\pmb{N}_{d,1}^T, \pmb{N}_{d,2}^T,  \dots,\pmb{N}_{d,K}^T]^T$. As mentioned, the quantization noise matrix $\pmb{Q}_{d,k} \in \mathbb{C}^{N_r \times T_d}$ is assumed to have i.i.d. $\mathcal{CN}(0,\sigma_{d,k}^2)$ entries, and hence the equivalent noise $\pmb{N}_{d,k}\in \mathbb{C}^{N_r \times T_d}$ the i.i.d. $\mathcal{CN}(0, P_d(\sigma_{z,k}^2+\sigma_{p,k}^2)/(P_pT_p)+\sigma_{z,k}^2+\sigma_{d,k}^2)$ entries. 

\subsection{Design Problem}\label{sec:pm_CDCS}

The objective of CDCS design is to maximize the ergodic achievable rate $R(\pmb{\sigma}_{p}^2,\pmb{\sigma}_{d}^2)$ for data communication subject to constraints on target detection performance and the uplink fronthaul rate $C_k^{\text{up}}(\sigma_{p,k}^2,\sigma_{d,k}^2)$. Optimization is over the fronthaul quantization noise power vectors $\pmb{\sigma}_{p}^2$ and $\pmb{\sigma}_{d}^2$, applied during pilot and data transmission, respectively. To this end, the problem is formulated as 
\begin{subequations}\label{eq:CDCS1}
\begin{eqnarray}
&&\hspace{-4cm} (\text{CDCS}:\pmb{P1})\hspace{0.5cm} \max_{\pmb{\sigma}_p^2, \pmb{\sigma}_d^2} \hspace{0.2cm} R(\pmb{\sigma}_{p}^2,\pmb{\sigma}_{d}^2) \label{eq:obj_CDCS1}\\
&&\hspace{-1cm} \text{s.t.} \hspace{0.5cm} \mathcal{B}(\pmb{\sigma}_p^2) \ge B_{\text{th}}, \label{eq:BConst_CDCS1}\\
&&\hspace{0cm}  C_k^{\text{up}}(\sigma_{p,k}^2,\sigma_{d,k}^2) \le \bar{C}_k, \hspace{0.2cm} k \in \mathcal{K},\label{eq:fConst_CDCS1}
\end{eqnarray}
\end{subequations}
where $B_{\text{th}}$ represents a target threshold for the radar sensing performance in terms of the Bhattacharyya distance $\mathcal{B}(\pmb{\sigma}_p^2)$. The Bhattacharyya distance provides an upper bound on the false alarm probability and a lower bound on the detection probability of sensing performance \cite{Kailath67TCOM}. In the following, we define the function $R(\pmb{\sigma}_{p}^2,\pmb{\sigma}_{d}^2)$, $\mathcal{B}(\pmb{\sigma}_p^2)$ and $C_k^{\text{up}}(\sigma_{p,k}^2,\sigma_{d,k}^2)$. 

For each AP $k \in \mathcal{K}$, the ergodic capacity of the CDCS strategy can be lower bounded as 
\begin{equation}\label{eq:CDCS_rate}
\frac{I(\pmb{x}_{d}; \hat{\pmb{Y}}_{d}|\tilde{\pmb{h}})}{T} \ge R(\pmb{\sigma}_{p}^2,\pmb{\sigma}_{d}^2), \hspace{0.2cm}\text{[bits/s/Hz]},
\end{equation} 
where $\pmb{\sigma}_p^2=\{\sigma_{p,k}^2\}_{k \in \mathcal{K}}$ and $\pmb{\sigma}_d^2=\{\sigma_{d,k}^2\}_{k \in \mathcal{K}}$, and we have defined 
\begin{equation}
R(\pmb{\sigma}_{p}^2,\pmb{\sigma}_{d}^2) = \frac{T_d}{T}E\left[\log_2\det\left(\pmb{I}_{N_rK}+P_d\tilde{\pmb{h}}\tilde{\pmb{h}}^H \pmb{\Omega}_N^{-1} \right)\right],
\end{equation}
with covariance matrices $\pmb{\Omega}_N = \text{diag} \{ \pmb{\Omega}_{N,1}, \dots,  \pmb{\Omega}_{N,K}\}$ and $\pmb{\Omega}_{N,k} = (P_d(\sigma_{z,k}^2+\sigma_{p,k}^2)/(P_pT_p)+\sigma_{z,k}^2+\sigma_{d,k}^2) \pmb{I}_{N_r}$. The bound (\ref{eq:CDCS_rate}) follows by treating the equivalent noise $\pmb{N}_{d,k}$ in (\ref{eq:Q_CDCSd}) as being zero-mean Gaussian independent with $\pmb{x}_d$ \cite{Kang14TWC, Hassibi03TIT}.

 The quantization noise powers $(\sigma_{p,k}^2, \sigma_{d,k}^2)$ for both training and data phase need to satisfy the fronthaul constraint $C_k(\sigma_{p,k}^2,\sigma_{d,k}^2) = C_{p,k}(\sigma_{p,k}^2) + C_{d,k}(\sigma_{d,k}^2) \le \bar{C}_k$, where 
\begin{subequations}\label{eq:Ck_CDCS_front}
\begin{eqnarray}
&& C_{p,k}(\sigma_{p,k}^2) = \frac{T_p}{T} E\left[\log_2\det\left(\pmb{I}_{N_r}+\frac{P_p\pmb{\Omega}_{h,k} + \sigma_{z,k}^2\pmb{I}_{N_r}}{\sigma_{p,k}^2}\right)\right]\label{eq:Ck_CDCS_frontp}\\
\text{and} && C_{d,k}(\sigma_{d,k}^2) = \frac{T_d}{T} E\left[\log_2\det\left(\pmb{I}_{N_r}+\frac{P_d\pmb{\Omega}_{h,k} + \sigma_{z,k}^2\pmb{I}_{N_r}}{\sigma_{d,k}^2}\right)\right],\label{eq:Ck_CDCS_frontd}
\end{eqnarray}
\end{subequations} 
respectively. In fact, by rate-distortion theory \cite{GamalBook}, the fronthaul rate $C_{d,k}(\sigma_{d,k}^2)$ in (\ref{eq:Ck_CDCS_frontd}) can be obtained via the bound 
\begin{subequations}\label{eq:Cd_CDCSd}
\begin{eqnarray}
C_{d,k}(\sigma_{d,k}^2) &=& \frac{1}{T} I(\pmb{Y}_{d,k}; \hat{\pmb{Y}}_{d,k})\\
&=& \frac{1}{T}\left(H(\pmb{Y}_{d,k}+\pmb{Q}_{d,k})-H(\pmb{Q}_{d,k})\right)\\
&\le& \frac{T_d}{T} E\left[\log_2\det\left(\pmb{I}_{N_r}+\frac{P_d\pmb{\Omega}_{h,k} + \sigma_{z,k}^2\pmb{I}_{N_r}}{\sigma_{d,k}^2}\right)\right],
\end{eqnarray}
\end{subequations} 
where the last inequality comes from the maximum entropy theorem \cite{Cov06}. Similar approach can be applied for $C_{p,k}(\sigma_{p,k}^2)$ of training phase. 

We further upper bound the sum-fronthaul requirement $C_k(\sigma_{p,k}^2,\sigma_{d,k}^2)$ by using Jensen's inequality \cite{Cov06} as
\begin{equation}\label{eq:CDCS_Jen}
C_k(\sigma_{p,k}^2,\sigma_{d,k}^2) \le C_k^{\text{up}}(\sigma_{p,k}^2,\sigma_{d,k}^2) = C_{p,k}^{\text{up}}(\sigma_{p,k}^2) + C_{d,k}^{\text{up}}(\sigma_{d,k}^2) 
\end{equation}
where we have defined $C_{p,k}^{\text{up}}(\sigma_{p,k}^2)=T_p/T \times \log_2\det(\pmb{I}_{N_r}+(P_pE[\pmb{\Omega}_{h,k}] + \sigma_{z,k}^2\pmb{I}_{N_r})/\sigma_{p,k}^2)$ and $C_{d,k}^{\text{up}}(\sigma_{d,k}^2)=T_d/T \times \log_2\det(\pmb{I}_{N_r}+(P_dE[\pmb{\Omega}_{h,k}] + \sigma_{z,k}^2\pmb{I}_{N_r})/\sigma_{d,k}^2)$. In (\ref{eq:CDCS_Jen}), we have
\begin{equation}\label{eq:CDCS_covh}
E[\pmb{\Omega}_{h,k}] = P_{\mathcal{H}_0}\sigma_{c,k}^2\pmb{I}_{N_r} + P_{\mathcal{H}_1}(\pmb{\Omega}_{g,k}+\sigma_{c,k}^2\pmb{I}_{N_r})=\sigma_{c,k}^2\pmb{I}_{N_r} + P_{\mathcal{H}_1}\pmb{\Omega}_{g,k},
\end{equation}
where $P_{\mathcal{H}_0}$ and $P_{\mathcal{H}_1}$ are the probabilities of target absence and existence, respectively, which satisfy the condition $P_{\mathcal{H}_0} + P_{\mathcal{H}_1}=1$. We assume that these probabilities are obtained based on the accumulated past history about the sensing results. 
  
For radar sensing, we adopt the Bhattacharyya distance to account for the detection performance of radar sensing, which measures the similarity of two discrete probability distributions \cite{Kailath67TCOM}. For zero-mean Gaussian distributions with covariance matrix of $\pmb{\Sigma}_1$ and $\pmb{\Sigma}_2$, the Bhattacharyya distance $\mathcal{B}$ is defined as \cite{Kailath67TCOM}
\begin{equation}\label{eq:Bdef}
\mathcal{B} = \log\left(\frac{\left|0.5\left(\pmb{\Sigma}_1+\pmb{\Sigma}_2\right)\right|}{\sqrt{|\pmb{\Sigma}_1||\pmb{\Sigma}_2|}}\right).
\end{equation} 
Therefore, based on the signal model (\ref{eq:HypoR_CDCSp}), the Bhattacharyya distance of two hypotheses in (\ref{eq:test_CDCSp}) can be calculated as 
\begin{eqnarray} 
\mathcal{B}(\pmb{\sigma}_p^2) &=&\sum_{k=1}^{K}\log\left(\frac{\left|\pmb{I}_{N_rT_p}+0.5\pmb{D}_k\pmb{\Lambda}_k\pmb{D}_k\right|}{\sqrt{\left|\pmb{I}_{N_rT_p}+\pmb{D}_k\pmb{\Lambda}_k\pmb{D}_k\right|}}\right)\nonumber\\
&=& \sum_{k=1}^{K}\log\det\left(\pmb{I}_{N_rT_p}+0.5\pmb{D}_k\pmb{\Lambda}_k\pmb{D}_k\right)- 0.5\log\det\left(\pmb{I}_{N_rT_p}+\pmb{D}_k\pmb{\Lambda}_k\pmb{D}_k\right)\nonumber
\end{eqnarray}
\begin{eqnarray}\label{eq:B_CDCS} 
&=& \sum_{k=1}^{K}\log\det\left(\pmb{I}_{N_rT_p}+\frac{0.5P_p\pmb{\Omega}_{g,k}}{P_p\sigma_{c,k}^2+\sigma_{z,k}^2+\sigma_{p,k}^2}\right) \nonumber\\
&&\hspace{4cm}- 0.5\log\det\left(\pmb{I}_{N_rT_p}+\frac{P_p\pmb{\Omega}_{g,k}}{P_p\sigma_{c,k}^2+\sigma_{z,k}^2+\sigma_{p,k}^2}\right).
\end{eqnarray}
Note that the Bhattacharyya distance $\mathcal{B}(\pmb{\sigma}_p^2)$ depends on the quantization noise variance for fronthaul link during training phase.

The problem $\pmb{P1}$ for CDCS method is non-convex due to the non-convex objective function and non-convex constraints (\ref{eq:BConst_CDCS1}) and (\ref{eq:fConst_CDCS1}). A heuristic algorithm based on line search is summarized in the appendix.

\section{Hybrid Cloud-based Decoding and Edge-based Sensing}\label{sec:CDES} 
In this section, we provide the problem formulation and the optimal fronthaul design for CDES. As seen in Fig. \ref{fig:fig_CDES}, in CDES, the APs detect the presence of the sensing target based on the received signal, then estimate the channel, and finally quantize the estimated channels and received signal. The CPU performs the channel decoding based on the quantized estimated channels during the pilot phase and received signals during the data phase, respectively.  

\subsection{Training Phase}\label{sec:CDESp}
In the CDES, the AP $k$ whitens the received signal in (\ref{eq:Hypo_p}) with the variances of the overall additive noise $\sqrt{P_p}\pmb{c}_k\pmb{s}_p^T + \pmb{Z}_{p,k}$, which leads to  
\begin{subequations}\label{eq:HypoR_CDESp}
\begin{eqnarray}
\text{(AP $k$)} && \mathcal{H}_0: \pmb{r}_{p,k} \sim \mathcal{CN}(\pmb{0}, \pmb{I}) \\
&& \mathcal{H}_1: \pmb{r}_{p,k} \sim \mathcal{CN}(\pmb{0}, \pmb{D}_k\pmb{\Lambda}_k\pmb{D}_k+\pmb{I}), 
\end{eqnarray}
\end{subequations}   
where $\pmb{r}_{p,k}$ and $\pmb{\Lambda}_k$ are equivalently defined in (\ref{eq:HypoR_CDCSp}), except defining $\pmb{D}_k$ as $\pmb{D}_k=\pmb{I}_{N_r}/\sqrt{P_p\sigma_{c,k}^2+\sigma_{z,k}^2}$. Therefore, the detection can be performed based on (\ref{eq:HypoR_CDESp}) by the following test 
\begin{align}\label{eq:test_CDESp}
        							&\mathcal{H}_1 \nonumber\\[-15pt]
\text{(AP $k$)} \hspace{0.2cm} \left(\pmb{r}_{p,k}\right)^H\pmb{T}_k\pmb{r}_{p,k}        &\gtreqless   \nu_p, \\[-15pt]
        							&\mathcal{H}_0 \nonumber
\end{align} 
where $\pmb{T}_k=\pmb{D}_k\pmb{\Lambda}_k\pmb{D}_k(\pmb{D}_k\pmb{\Lambda}_k\pmb{D}_k+\pmb{I}_{N_rT_p})^{-1}$. In the case of edge-based sensing, the AP $k$ transmits the obtained $1$-bit hard decision to the CPU. Accordingly, the CDES method requires $1/T$ bits/sample for transferring the radar sensing results to the CPU. Then, the CPU decides on the target's presence by the majority rule. In other words, if the number $N_R$ of APs to determine $\mathcal{H}_0$ satisfies $N_R \ge K/2$, the CPU chooses $\mathcal{H}_0$, and vice versa if $N_R \le K/2$.    

Based on (\ref{eq:Sigp}), the channels are estimated at each AP $k$ by the ML method as in (\ref{eq:ch_CDCSp}). Similar to CDCS, the received signal in (\ref{eq:Sigp}), the ML estimate of $\pmb{h}_k$ is given as  
\begin{equation}\label{eq:ch_CDESp}
\text{(AP $k$)} \hspace{0.2cm} \tilde{\pmb{h}}_k = \pmb{Y}_{p,k}\frac{\pmb{x}_p^*}{\|\pmb{x}_p\|^2}= \pmb{h}_k + \pmb{Z}_{p,k}\frac{\pmb{x}_p^*}{\|\pmb{x}_p\|^2} = \pmb{h}_k + \pmb{e}_k, 
\end{equation}
for all $k$, where the estimation error $\pmb{e}_{k} \in \mathbb{C}^{N_r \times 1}=\pmb{Z}_{p,k}\frac{\pmb{x}_p^*}{\|\pmb{x}_p\|^2}$ has i.i.d. $\mathcal{CN}(0, \sigma_{z,k}^2/(P_pT_p))$. Similar to CDCS, the variance $\pmb{\Omega}_{\tilde{h},k|\mathcal{H}_i}$ of the channel estimate $\tilde{\pmb{h}}_k$ is varied by the hypothesis $\mathcal{H}_i$ of the target presence, for $i \in \{0,1\}$, and we have    
\begin{subequations}\label{eq:ML_var_CDES}
\begin{eqnarray}
&& \hspace{-0.8cm}\mathcal{H}_0: \pmb{\Omega}_{\tilde{h},k|\mathcal{H}_0} = \left(\frac{\sigma_{c,k}^2P_pT_p-\sigma_{z,k}^2}{P_pT_p}\right)\pmb{I},\\
&& \hspace{-0.8cm} \mathcal{H}_1: \pmb{\Omega}_{\tilde{h},k|\mathcal{H}_1} = \pmb{\Omega}_{g,k}+\left(\frac{\sigma_{c,k}^2P_pT_p-\sigma_{z,k}^2}{P_pT_p}\right)\pmb{I}.
\end{eqnarray}
\end{subequations}  

Unlike CDCS, since the channel is estimated at each AP $k$, the variance of the channel estimate is not related to the fronthaul design of pilot phase. After that, the channel estimate $\tilde{h}_k$ at the AP $k$ is compressed and forwarded to the CPU on the fronthaul link, and then the compressed channel is given as
\begin{equation}\label{eq:hath_CDES}
\tilde{\pmb{h}}_k= \hat{\pmb{h}}_k + \pmb{q}_{p,k}, 
\end{equation}
for all $k$, where the $N_r \times 1$ quantization noise vector $\pmb{q}_{p,k}$ has the zero-mean i.i.d $\mathcal{CN}(0, \sigma_{p,k}^2)$ entries; and the compressed estimate $\hat{\pmb{h}}_k$ is the complex Gaussian with zero mean and covariance matrix $\pmb{\Omega}_{\tilde{h},k|\mathcal{H}_i} - \sigma_{p,k}^2\pmb{I}_{N_r}$ for $i \in \{0,1\}$ based on the hypothesis $\mathcal{H}_1$ or $\mathcal{H}_0$ depending on the sensing result. Note that we consider a different model for the quantization test channel (see, e.g., \cite[Ch. 3]{GamalBook}) in (\ref{eq:hath_CDES}) compared to (\ref{eq:Q_CDCSd}). In (\ref{eq:hath_CDES}), the quantization noise is added to the compressed signal, while the approach of (\ref{eq:Q_CDCSd}) is optimal according to the rate-distortion theory \cite[Ch. 3]{GamalBook}. Therefore, the test channel (\ref{eq:hath_CDES}) is adopted here for its analytical convenience, which is assumed in many existing studies, e.g., \cite{San09TIT, Lim11TIT}. We will discuss the quantization noise $\sigma_{p,k}^2$ to the fronthaul capacity $\bar{C}_k$.
        					
\subsection{Data Phase}\label{sec:CDESd} 
In CDES, each AP $k$ compresses and transfers the $N_r \times T_d$ received data signal $\pmb{Y}_{d,k}$ in (\ref{eq:Sigd}) over the fronthaul capacity. Accordingly, the received signal at the CPU is given as
\begin{equation}\label{eq:Q_CDESd}
\hat{\pmb{Y}}_{d} = \pmb{Y}_{d} + \pmb{Q}_{d} = \hat{\pmb{h}}\pmb{x}_d^T + \pmb{N}_{d},
\end{equation}
where we have defined $\hat{\pmb{Y}}_{d} = [\hat{\pmb{Y}}_{d,1}^T, \hat{\pmb{Y}}_{d,2}^T, \dots, \hat{\pmb{Y}}_{d,K}^T]^T$ with $\hat{\pmb{Y}}_{d,k} = \hat{\pmb{h}}_k\pmb{x}_d^T + \pmb{N}_{d,k}$, $\hat{\pmb{h}} = [\hat{\pmb{h}}_1^T, \hat{\pmb{h}}_2^T, \dots, \hat{\pmb{h}}_K^T]^T$, $\pmb{Q}_{d} = [\pmb{Q}_{d,1}^T, \pmb{Q}_{d,2}^T,$ $\dots,\pmb{Q}_{d,K}^T]^T$ with the quantization noise $\pmb{Q}_{d,k} \in \mathbb{C}^{N_r \times T_d}$ to have i.i.d. $\mathcal{CN}(0,\sigma_{d,k}^2)$ entries and $\pmb{N}_{d} = [\pmb{N}_{d,1}^T, \pmb{N}_{d,2}^T,$ $\dots,\pmb{N}_{d,K}^T]^T$ with the equivalent noise being $\pmb{N}_{d,k}=(\pmb{q}_{p,k}+\pmb{e}_k)\pmb{x}_d^T+\pmb{Z}_{d,k}+\pmb{Q}_{d,k}$ to have i.i.d. $\mathcal{CN}(0, P_d(\sigma_{p,k}^2+\sigma_{z,k}^2/(P_pT_p))+\sigma_{z,k}^2+\sigma_{d,k}^2)$ entries.

\subsection{Design Problem}\label{sec:pm_CDES}
As done in the previous section, we formulate the design problem as the optimization of the ergodic achievable rate $R(\pmb{\sigma}_{p}^2,\pmb{\sigma}_{d}^2)$ of data communication subject to fronthaul and sensing constraints. 
Accordingly, the problem is formulated as 
\begin{subequations}\label{eq:CDES1}
\begin{eqnarray}
&&\hspace{-4cm} (\text{CDES}:\pmb{P2})\hspace{0.5cm} \max_{\pmb{\sigma}_p^2, \pmb{\sigma}_d^2} \hspace{0.2cm} R(\pmb{\sigma}_{p}^2,\pmb{\sigma}_{d}^2) \label{eq:obj_CDES1}\\
&&\hspace{-1cm} \text{s.t.} \hspace{0.5cm} \frac{1}{T} + C_k^{\text{up}}(\sigma_{p,k}^2,\sigma_{d,k}^2) \le \bar{C}_k, \hspace{0.2cm} k \in \mathcal{K},\label{eq:fConst_CDES1}
\end{eqnarray}
\end{subequations}
where we recall that $1/T$ is the fronthaul overhead for sensing. In the following, we define the function $R(\pmb{\sigma}_{p}^2,\pmb{\sigma}_{d}^2)$, $\mathcal{B}(\pmb{\sigma}_p^2)$ and $C_k^{\text{up}}(\sigma_{p,k}^2,\sigma_{d,k}^2)$ in (\ref{eq:CDES1}) for CDES. 

For each AP $k \in \mathcal{K}$, the ergodic capacity of the CDES is given as  
\begin{equation}\label{eq:Rk_CDES}
\frac{I(\pmb{x}_{d}; \hat{\pmb{Y}}_{d}|\hat{\pmb{h}})}{T} \ge  R(\pmb{\sigma}_{p}^2,\pmb{\sigma}_{d}^2),
\end{equation} 
with respect to the distribution of the target existence, where we define 
\begin{equation}
 R(\pmb{\sigma}_{p}^2,\pmb{\sigma}_{d}^2) = \frac{T_d}{T}E\left[\log_2\det\left(\pmb{I}_{N_r K}+ P_d\hat{\pmb{h}}\hat{\pmb{h}}^H \pmb{\Omega}_N^{-1} \right)\right].
\end{equation}
where $\pmb{\Omega}_N = \text{diag} \{ \pmb{\Omega}_{N,1}, \dots,  \pmb{\Omega}_{N,K}\}$ with $\pmb{\Omega}_{N,k} = (P_d(\sigma_{p,k}^2+\sigma_{z,k}^2/(P_pT_p))+\sigma_{z,k}^2+\sigma_{d,k}^2) \pmb{I}_{N_r}$. 

The quantization noise powers $(\sigma_{p,k}^2, \sigma_{d,k}^2)$ for both training and data phase need to satisfy the ergodic fronthaul constraint $C_k(\sigma_{p,k}^2,\sigma_{d,k}^2) = C_{p,k}(\sigma_{p,k}^2) + C_{d,k}(\sigma_{d,k}^2) \le \bar{C}_k - 1/T$, where 
\begin{subequations}\label{eq:Ck_CDESd}
\begin{eqnarray}
&& C_{p,k}(\sigma_{p,k}^2) =  \frac{1}{T} E\left[ \log_2\det \left(\pmb{I}_{N_r}+\frac{\pmb{\Omega}_{\hat{ h},k}}{\sigma_{p,k}^2}\right)\right] \\
\text{and} && C_{d,k}(\sigma_{d,k}^2) = \frac{T_d}{T} E\left[ \log_2\det\left(\pmb{I}_{N_r}+\frac{P_d\pmb{\Omega}_{h,k} + \sigma_{z,k}^2\pmb{I}_{N_r}}{\sigma_{d,k}^2}\right) \right],
\end{eqnarray}
\end{subequations} 
respectively, where $\pmb{\Omega}_{\hat h,k}$ depends on the sensing results about the target existence. In a similar manner for (\ref{eq:Cd_CDCSd}), the fronthaul rate $C_{p,k}(\sigma_{p,k}^2)$ for training phase can be derived as 
\begin{subequations}\label{eq:Cp_CDESd}
\begin{eqnarray}
C_{p,k}(\sigma_{p,k}^2) &=& \frac{1}{T} I(\tilde{\pmb{h}}_{k}; \hat{\pmb{h}}_{k}) = \frac{1}{T}\left(H(\hat{\pmb{h}}_{k}+\pmb{q}_{p,k})-H({\pmb{q}}_{p,k})\right)\\
&\le& \frac{1}{T} E\left[ \log_2\det \left(\pmb{I}_{N_r}+\frac{\pmb{\Omega}_{\hat{ h},k}}{\sigma_{p,k}^2}\right)\right],
\end{eqnarray}
\end{subequations} 
This is done by treating the equivalent noise $\pmb{q}_{p,k}$ in (\ref{eq:hath_CDES}) and $\pmb{Q}_{d,k}$ in (\ref{eq:Q_CDESd}) as being zero-mean Gaussian independent with $\pmb{h}_k$ and $\pmb{x}_d$ \cite{Kang14TWC, Hassibi03TIT}. By rate-distortion theory \cite{GamalBook}, we can derive them similarly with (\ref{eq:Cd_CDCSd}). As in (\ref{eq:CDCS_Jen}), the ergodic fronthaul capacity can be upper-bounded by using Jensen\rq{s} inequality \cite{Ross22Book} as 
\begin{equation}\label{eq:CDES_Jen}
C_k(\sigma_{p,k}^2,\sigma_{d,k}^2) \le C_k^{\text{up}}(\sigma_{p,k}^2,\sigma_{d,k}^2) = C_{p,k}^{\text{up}}(\sigma_{p,k}^2) + C_{d,k}^{\text{up}}(\sigma_{d,k}^2) 
\end{equation}
where we have defined $C_{p,k}^{\text{up}}(\sigma_{p,k}^2)=1/T \times \log_2\det(\pmb{I}_{N_r}+E[\pmb{\Omega}_{\hat{h},k}]/\sigma_{p,k}^2)$ and $C_{d,k}^{\text{up}}(\sigma_{d,k}^2)=T_d/T \times \log_2\det(\pmb{I}_{N_r}+(P_dE[\pmb{\Omega}_{h,k}] + \sigma_{z,k}^2\pmb{I}_{N_r})/\sigma_{d,k}^2)$ as the upperbound of the ergodic fronthaul rates for the pilot phase and data phase, respectively. In (\ref{eq:CDES_Jen}), we note that the expected value of $E[\pmb{\Omega}_{\tilde h,k}] = P_{\mathcal{H}_0} \pmb{\Omega}_{\tilde{h},k|\mathcal{H}_0} + P_{\mathcal{H}_1}\pmb{\Omega}_{\tilde{h},k|\mathcal{H}_1}=\left(\frac{\sigma_{c,k}^2P_pT_p-\sigma_{z,k}^2}{P_pT_p}\right)\pmb{I}_{N_r} + P_{\mathcal{H}_1}\pmb{\Omega}_{g,k}$, where $P_{\mathcal{H}_0}$ and $P_{\mathcal{H}_1}$ are the probabilities of target absence and existence, respectively, which satisfy the condition $P_{\mathcal{H}_0} + P_{\mathcal{H}_1}=1$. 

The problem $\pmb{P2}$ for CDES is non-convex due to the non-convex objective function and non-convex constraint. To this end, we apply a heuristic algorithm based on line search, where the maximal value of the ergodic achievable rate (\ref{eq:obj_CDES1}) can be obtained when the equality holds in (\ref{eq:fConst_CDES1}), i.e., the fronthaul capacity is fully consumed for both pilot and data phases. This follows in a manner similar to Algorithm \ref{alg:CDCS} in the appendix. 

\section{Hybrid Edge-based Decoding and Cloud-based Sensing}\label{sec:EDCS} 
In EDCS, as seen in Fig. \ref{fig:fig_EDCS}, the APs estimate the channel, and then decode the communication signal. The channel estimate is refined based on the decoded data. The channel is finally forwarded to the CPU from all APs for radar sensing, but the decoded data is forwarded from only one AP with the highest SNR for efficient usage of fronthaul link, whose details are provided in the following.

\subsection{Training Phase}\label{sec:EDCSp} 
In the EDCS, since each AP estimates the channel and decodes the data signal, we adopt the ML method for channel estimation at APs as in CDES scheme. Therefore, the ML estimate of $\pmb{h}_k$ is equivalent to that of CDES in (\ref{eq:ch_CDESp}).  

\subsection{Data Phase}\label{sec:EDCSd}
Based on the channel estimates $\tilde{\pmb{h}}_k$ in (\ref{eq:ch_CDESp}) during the training phase, the AP $k$ decodes the $N_r \times T_d$ $\pmb{Y}_{d,k}$, and refines the channel estimates $\tilde{\pmb{h}}_k$ by using the decoded message. In particular, it is noted that since $\pmb{x}_d$ is an encoded codeword, each AP $k$ applies channel decoding to recover the codeword $\pmb{x}_d$. By following \cite{Cov06}, if the transmission rate is chosen to be below the corresponding maximum achievable rate, then the APs can correctly decode $\pmb{x}_d$ with the arbitrarily high probability as the block-length increases, that is, we can have $\tilde{\pmb{x}}_d = \pmb{x}_d$, where $\tilde{\pmb{x}}_d$ is the decoded codeword at APs. 

With the decoded data $\tilde{\pmb{x}}_d$, each AP refines the channel estimate as follows. To transfer the channel estimate to the CPU for radar sensing, the AP $k$ collects and redefines the received signals and noise signals of both training and data phase as $\pmb{Y}_k = [\pmb{Y}_{p,k} \,\,\, \pmb{Y}_{d,k}] \in \mathbb{C}^{N_r \times T}$ and $\pmb{Z}_k = [\pmb{Z}_{p,k} \,\,\, \pmb{Z}_{d,k}] \in \mathbb{C}^{N_r \times T}$, respectively. Then, we can rewrite the received signal at AP $k$ during the $T$ channel uses as 
\begin{equation}\label{eq:rx_EDCS}
\pmb{Y}_k = \pmb{h}_k\pmb{x}^T + \pmb{Z}_k = \pmb{h}_k\tilde{\pmb{x}}^T + \pmb{Z}_k,
\end{equation}   
where we define $\tilde{\pmb{x}}=[\pmb{x}_p ; \tilde{\pmb{x}}_d] \in \mathbb{C}^{T \times 1}$ with $\pmb{x}=[\pmb{x}_p; \pmb{x}_d] \in \mathbb{C}^{T \times 1}$. By using the ML method, the refined channel estimate is expressed as
\begin{equation}\label{eq:ch_EDCStot}
\bar{\pmb{h}}_k = \pmb{Y}_k\frac{\tilde{\pmb{x}}^*}{\|\tilde{\pmb{x}}\|^2} = \pmb{h}_k + \pmb{Z}_{k}\frac{\tilde{\pmb{x}}^*}{\|\tilde{\pmb{x}}\|^2} = \pmb{h}_k + \bar{\pmb{e}}_k,
\end{equation}
where $\bar{\pmb{e}}_{k}\in \mathbb{C}^{N_r \times 1}=\pmb{Z}_{k}\frac{\tilde{\pmb{x}}^*}{\|\tilde{\pmb{x}}\|^2}$ represents the estimation error with i.i.d. $\mathcal{CN}(0, \sigma_{z,k}^2/(P_pT_p+P_dT_d))$.

For radar sensing, similar with (\ref{eq:hath_CDES}), the AP $k$ transfers the channel estimates $\bar{\pmb{h}}_k$ to the CPU, and then the received signal $\hat{\pmb{h}}_k$ at the CPU can be written as
\begin{equation}\label{eq:ch_EDCSd_CU}
\text{(CPU)} \hspace{0.2cm} \bar{\pmb{h}}_k = \hat{\pmb{h}}_k + \pmb{q}_k, 
\end{equation}
where $\pmb{q}_k \in \mathcal{C}^{N_r \times 1}$ is the quantization noise vector with i.i.d $\mathcal{CN}(0, \sigma_k^2)$ entries. The CPU collects all received signals from APs, and then whitens them with the variance of $\sigma_k^2$, which are given as  
\begin{subequations}\label{eq:Hypoh_EDCS}
\begin{eqnarray}
\text{(CPU)} && \mathcal{H}_0: \hat{\pmb{h}}^W \sim \mathcal{CN}(\pmb{0}, \pmb{I}) \\
&& \mathcal{H}_1: \hat{\pmb{h}}^W \sim \mathcal{CN}(\pmb{0}, \pmb{D}\pmb{\Lambda}\pmb{D}+\pmb{I}), 
\end{eqnarray}
\end{subequations}   
where $\hat{\pmb{h}}^W = [(\hat{\pmb{h}}^W_1)^T \cdots (\hat{\pmb{h}}^W_K)^T]^T$ with $\hat{\pmb{h}}^W_k = \pmb{D}_{k}\hat{\pmb{h}}_k$; $\pmb{D}_k$ is the whitening matrix for AP $k$ given by $\pmb{D}_k=\sqrt{P_pT_p+P_dT_d}\pmb{I}_{N_r}/\sqrt{(P_pT_p+P_dT_d)(\sigma_{c,k}^2-\sigma_{k}^2) + \sigma_{z,k}^2}$; $\pmb{D}=\text{diag}\{\pmb{D}_1, \dots, \pmb{D}_{K}\}$; and $\pmb{\Lambda} = \text{diag}\{\pmb{\Omega}_{g,1}, \dots, \pmb{\Omega}_{g,K}\}$. The detection problem in (\ref{eq:Hypoh_EDCS}) has the standard Neyman-Pearson solution obtained by the test 
\begin{align}\label{eq:test_EDCS}
        							&\mathcal{H}_1 \nonumber\\[-15pt]
\text{(CPU)} \hspace{0.2cm} (\hat{\pmb{h}}^W)^H\pmb{T}\hat{\pmb{h}}^W        &\gtreqless   \nu_p, \\[-15pt]
        							&\mathcal{H}_0 \nonumber
\end{align} 
where we have defined $\pmb{T}=\pmb{D}\pmb{\Lambda}\pmb{D}(\pmb{D}\pmb{\Lambda}\pmb{D}+\pmb{I}_{N_r})^{-1}$, and $\nu_p$ is the threshold based on the tolerated false alarm probability \cite{Kay93Book}. 

\subsection{Design Problem}\label{sec:pm_EDCS}
Recalling that AP with the highest SNR carries out decoding and denoting the index of AP with the highest SNR as $k=1$, the optimization problem of interest can be formulated as
\begin{subequations}\label{eq:EDCS1}
\begin{eqnarray}
(\text{EDCS}:\pmb{P3})  && \max_{\pmb{\sigma}^2, R_1} \hspace{0.2cm} R_1 \label{eq:obj_EDCS1}\\
&& \hspace{0.3cm} \text{s.t.} \hspace{0.5cm} \mathcal{B}(\pmb{\sigma}^2) \ge B_{\text{th}}, \label{eq:fConst_EDCS1} \\
&&\hspace{0.8cm} \hspace{0.5cm} R_1 \le \frac{T_d}{T}E\left[\log_2\det\left(\pmb{I}_{N_r}+\frac{P_pP_dT_p}{\sigma_{z, 1}^2\left(P_d+P_pT_p\right)}\tilde{\pmb{h}}_{1}\tilde{\pmb{h}}_{1}^H\right)\right], \label{eq:fConst_EDCS2} \\
&&\hspace{0.8cm} \hspace{0.5cm} C_{1}^{\text{up}}(\sigma_{1}^2) \le \bar{C}_1 - R_1, \label{eq:fConst_EDCS3} \\
&&\hspace{0.8cm} \hspace{0.5cm} C_{k}^{\text{up}}(\sigma_{k}^2) \le \bar{C}_k, \hspace{0.2cm} k \in \mathcal{K} \backslash 1, \label{eq:fConst_EDCS4}
\end{eqnarray}
\end{subequations}
where $\pmb{\sigma}^2=\{\sigma_k^2\}_{\forall k}$ and we have defined the set excluding the AP $1$ as $\mathcal{K} \backslash 1 = \mathcal{K}-\{1\}$.
       
The ergodic capacity of AP can be derived as the mutual information $I(\pmb{X}_{d,1};\pmb{Y}_{d,1}|\tilde{\pmb{h}}_1)/T$ [bits/s/Hz] is bounded as 
\begin{equation}\label{eq:Rk_EDCS}
\frac{I(\pmb{X}_{d,1}; \pmb{Y}_{d,1}|\tilde{\pmb{h}}_1)}{T} \ge R_{1} = \frac{T_d}{T}E\left[\log_2\det\left(\pmb{I}_{N_r}+\frac{P_pP_dT_p}{\sigma_{z,1}^2\left(P_d+P_pT_p\right)}\tilde{\pmb{h}}_1\tilde{\pmb{h}}_1^H\right)\right],
\end{equation} 
where $\tilde{\pmb{h}}_1$ is defined as in (\ref{eq:ch_CDESp}), and $\pmb{Y}_{d,1}$ is the received signal at AP $1$ in data phase. The quantization noise power $\sigma_{k}^2$ needs to satisfy the fronthaul constraint, where 
\begin{equation} \label{eq:Ck_EDCSd}
C_{k}(\sigma_{k}^2) =  \frac{1}{T} I(\bar{\pmb{h}}_{k}; \hat{\pmb{h}}_{k}) \le \frac{1}{T} E \left[\log_2\det\left(\pmb{I}_{N_r}+ \frac{\pmb{\Omega}_{\hat{ h},k}}{\sigma_{k}^2}\right) \right ].
\end{equation}
As in (\ref{eq:CDES_Jen}), by using Jensen\rq{s} inequality \cite{Ross22Book}, an upperbound on (\ref{eq:Ck_EDCSd}) is calculated as 
\begin{equation} \label{eq:EDCS_Jen}
C_{k}(\sigma_{k}^2)  \le C_{k}^{\text{up}}(\sigma_{k}^2)= \frac{1}{T}  \log_2\det\left(\pmb{I}_{N_r}+\frac{E \left[\pmb{\Omega}_{\hat{ h},k} \right ]}{\sigma_{k}^2}\right),
\end{equation}
where $E[\pmb{\Omega}_{\hat{h},k}] = P_{\mathcal{H}_0} \pmb{\Omega}_{\hat{h},k|\mathcal{H}_0} + P_{\mathcal{H}_1}\pmb{\Omega}_{\hat{h},k|\mathcal{H}_1}= (\sigma_{c,k}^2-\sigma_{k}^2 +\sigma_{z,k}^2/(P_pT_p + P_d T_d))\pmb{I}_{N_r} + P_{\mathcal{H}_1}\pmb{\Omega}_{g,k}$ with the probability of target absence $P_{\mathcal{H}_0}$ and the probability of target existence $P_{\mathcal{H}_1}$. For the AP $1$ with the highest SNR, under the fronthaul capacity constraint, the fronthaul capacity to be used to convey the channel estimate is $\bar{C}_1 - R_1$, where $R_1$ is upper-bounded by (\ref{eq:Rk_EDCS}) when $k=1$. 

Based on (\ref{eq:Hypoh_EDCS}), the Bhattacharyya distance of two hypotheses in (\ref{eq:test_EDCS}) can be calculated as 
\begin{equation}\label{eq:B_EDCS}
\mathcal{B}(\pmb{\sigma}^2) =\sum_{k=1}^{K}\log\det\left(\pmb{I}_{N_r}+0.5\lambda_k\pmb{\Omega}_{g,k}\right) -0.5\log\det\left(\pmb{I}_{N_r}+\lambda_k\pmb{\Omega}_{g,k}\right),
\end{equation}         
where we have defined 
\begin{equation}\label{eq:lambda_EDCS}
\lambda_k=\frac{P_pT_p+P_dT_d}{\left(P_pT_p+P_dT_d\right)(\sigma_{c,k}^2- \sigma_{k}^2)+\sigma_{z,k}^2}.
\end{equation} 
Note that, as in CDCS, the Bhattacharyya distance $\mathcal{B}(\pmb{\sigma}_k^2)$ depends on the quantization noise variance for fronthaul link. The solution of (\ref{eq:EDCS1}) can be obtained by a heuristic algorithm based on line search, where the maximal value of the ergodic achievable rate (\ref{eq:obj_EDCS1}) is obtained when the equality holds in (\ref{eq:fConst_EDCS2}), (\ref{eq:fConst_EDCS3}) and (\ref{eq:fConst_EDCS4}), i.e., the fronthaul capacity is fully consumed for all APs. This again follows the approach in the appendix.   

\section{Edge-based Decoding and Edge-based Sensing}\label{sec:EDES}
In this section, we finally explore EDES, whereby, as shown in Fig. \ref{fig:fig_EDES}, each AP performs both decoding and sensing individually. Based on the channel estimates and radar sensing results, the APs decode the data signals during the data phase. Note that the EDES method does not require the fronthaul links. 

\subsection{Training Phase}\label{sec:EDESp}
In the EDES, the AP $k$ whitens the received signal as in (\ref{eq:HypoR_CDESp}) of the CDES case, and then performs the radar sensing based on (\ref{eq:HypoR_CDESp}) by the test (\ref{eq:test_CDESp}). By using the ML method as $\tilde{\pmb{h}}_k$ like CDES, the channels can be estimated at each AP $k$, where the variances of the estimated channel and the estimation error under the hypothesis $\mathcal{H}_0$ or $\mathcal{H}_1$ are equivalent with those of (\ref{eq:ch_CDESp}). 

\subsection{Data Phase}\label{sec:EDESd}
After the channel estimation, during the data phase, the AP $k$ decodes the $N_r \times T_d$ received signal $\pmb{Y}_{d,k}$ based on the channel estimates obtained during the training phase. Based on the channel estimates $\tilde{\pmb{h}}_k$ in (\ref{eq:ch_CDESp}) during the training phase, the AP $k$ decodes the $N_r \times T_d$ received signal $\pmb{Y}_{d,k}$. Consequently, the received signal $\pmb{Y}_{d,k}$ can be written with the equivalent noise $\pmb{N}_{d,k}=\pmb{x}_d\pmb{e}_k^T+\pmb{Z}_{d,k}^T$ as 
\begin{equation}\label{eq:Y_EDESd}
\pmb{Y}_{d,k} = \tilde{\pmb{h}}_k\pmb{x}_d^T + \pmb{N}_{d,k},
\end{equation}
where the equivalent noise $\pmb{N}_{d,k}$ has a zero-mean and covariance matrix $\sigma_{z,k}^2(P_d/(P_pT_p)+1)\pmb{I}_{N_r}$.

\subsection{Design Problem}\label{sec:pm_EDES}
In EDES, one AP $k=1$ with the highest SNR carries out decoding like EDCS, while the target's presence is determined via majority rule at CPU based on the forwarded sensing results from all APs as in CDES. Accordingly, the user rate $R_1$ is designed and then used over the fronthaul link for the AP $1$, and the $1$ bit for the sensing result is conveyed over the fronthaul links from all APs. Accordingly, the optimization problem of EDES can be formulated as
\begin{subequations}\label{eq:EDES1}
\begin{eqnarray}
(\text{EDES}:\pmb{P4})  && \max_{R_1} \hspace{0.2cm} R_1 \label{eq:obj_EDES1}\\
&& \hspace{0.3cm} \text{s.t.} \hspace{0.5cm} R_1 \le \frac{T_d}{T}E\left[\log_2\det\left(\pmb{I}_{N_r}+\frac{P_pP_dT_p}{\sigma_{z, 1}^2\left(P_d+P_pT_p\right)}\tilde{\pmb{h}}_1\tilde{\pmb{h}}_1^H\right)\right], \label{eq:fConst_EDES1} \\
&&\hspace{0.8cm} \hspace{0.5cm} R_1 + \frac{1}{T} \le \bar{C}_1, \label{eq:fConst_EDES2} \\
&&\hspace{0.8cm} \hspace{0.5cm} \frac{1}{T} \le \bar{C}_k, \hspace{0.2cm} k \in \mathcal{K} \backslash 1. \label{eq:fConst_EDES3}
\end{eqnarray}
\end{subequations}

Note that if fronthaul link capacity is not extremely small, i.e., $\frac{1}{T} \le \bar{C}_k$ for $k \in \mathcal{K}$, the rate $R_1$ can be evaluated as 
\begin{equation}
R_1 = \min \left ( \frac{T_d}{T}E\left[\log_2\det\left(\pmb{I}_{N_r}+\frac{P_pP_dT_p}{\sigma_{z,1}^2\left(P_d+P_pT_p\right)}\tilde{\pmb{h}}_{1}\tilde{\pmb{h}}_{1}^H\right)\right], \bar C_1 - \frac{1}{T}  \right ).
\end{equation}

\section{Numerical Results}\label{sec:num}
In this section, the performance of the considered PMN schemes, namely CDCS, CDES, EDCS and EDES, is investigated via numerical results. Throughout, we consider $K$ APs with $N_r=2$ antennas, which are uniformly located in $[0, 100]$ on the x-axis, while a sensing target and a single UE are located at $(20, 50)$ and $(50, 50)$, respectively. Moreover, we assume that the transmit powers for training and data transmission phases need to satisfy the same power constraint of $P_T$, and each AP has the same fronthaul capacity $\bar C$, that is, $P_p = P_d = P_T$ and $\bar C_k = \bar C$, for $k \in \mathcal{K}$. We set the variance of the communication and sensing channel gain as $\sigma_{\alpha,k}^2 = 0.1$, the variance of the clutter components as $\sigma_{c,k}^2 = 0.01$, for all $k \in \mathcal{K}$, and the variance of the noise as $\sigma_{z,k}^2 = 1$ for all $k \in \mathcal{K}$ \cite{JSA16ETT}. Also, we set the duration of pilot phase and total phase as $T_p = N_r$ and $T=10$, respectively. 

\begin{figure}[t]
\centering
\includegraphics[width=12cm]{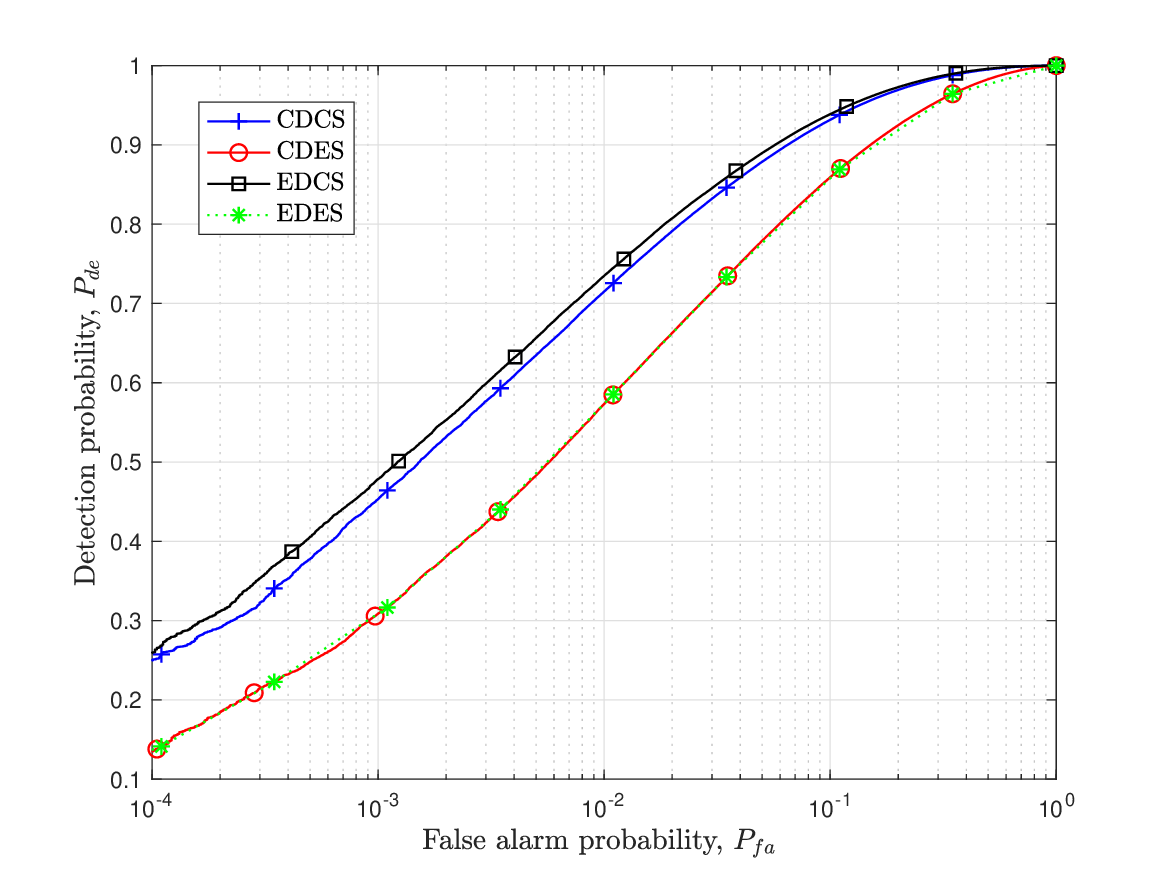}
\caption{ROC curves with different functional splits ($B_{th}=6$, $K=7$, $N_r=2$, $P_T=23$ dB and $\bar{C}=10$)}
\label{fig:ROC}
\end{figure}

\begin{figure}[t]
\centering
\includegraphics[width=12cm]{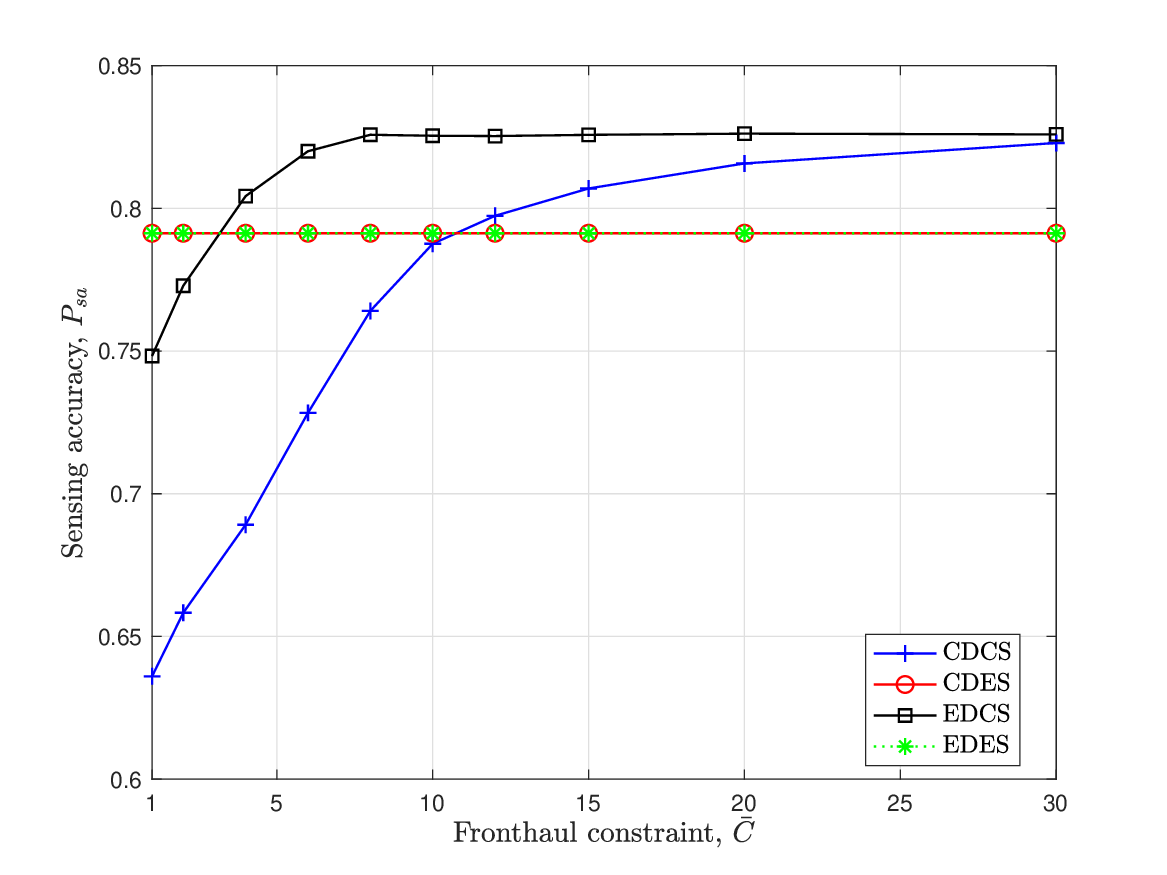}
\caption{Sensing accuracy versus fronthaul capacity constraint $\bar{C}$ ($K=3$, $B_{th}=2$, and $P_T=23$ dB)}
\label{fig:Prob_sensing_vsC}
\end{figure}

\begin{figure}[h!]
\centering
\includegraphics[width=12cm]{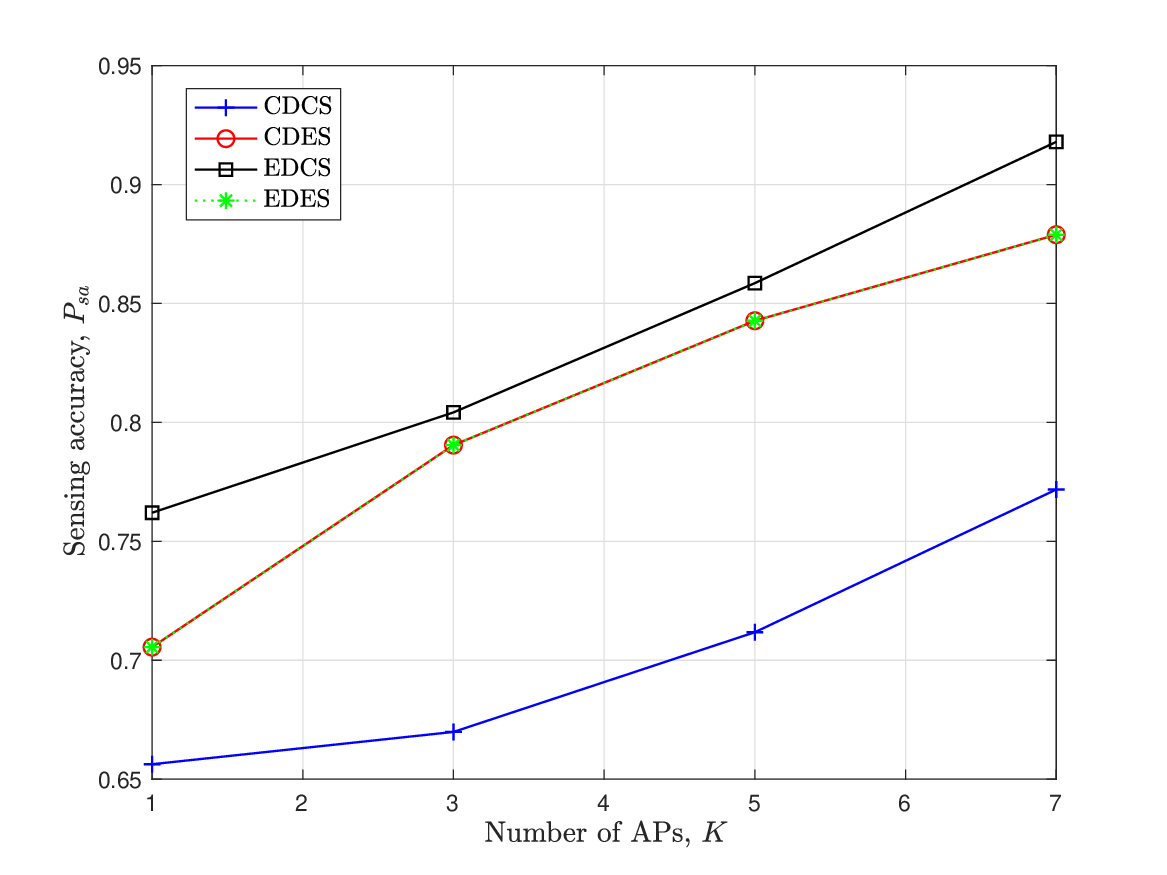}
\caption{Sensing accuracy versus the number $K$ of APs  ($\bar C=4$, $B_{th}=2$, and $P_T=23$ dB)}
\label{fig:Prob_sensing_vsK}
\end{figure}

As the performance criteria, we focus on the receiving operating characteristic (ROC) curves and the sensing accuracy as well as on the ergodic achievable rate. Here, the sensing accuracy $P_{sa}$ is defined as the average of the probability of the true positive and the true negative (correct rejection), i.e., 
\begin{equation}\label{eq:sa}
P_{sa} = \frac{1}{2} (P_{de} + (1-P_{fa})),
\end{equation}
where $P_{de}$ and $P_{fa}$ represent the detection probability and the false alarm probability, respectively. 

\subsection{Cloud vs. Edge Sensing}\label{sec:CE_sensing}

We start by evaluating the performance in terms of sensing. In Fig. \ref{fig:ROC}, the ROC curves, i.e., the detection probability $P_{de}$ versus the false alarm probability $P_{fa}$, of the functional splits are illustrated. The ROC curves are obtained by implementing the optimum test detectors (\ref{eq:test_CDCSp}), (\ref{eq:test_CDESp}) and (\ref{eq:test_EDCS}) and varying the decision threshold $\nu_p$. It is observed that the cloud-based sensing (CDCS and EDCS) outperforms edge-based sensing (CDES and EDES), for the given fronthaul capacity. Furthermore, EDCS improves over CDCS, since it devotes the fronthaul capacity exclusively for sensing. It is also noted that CDES and EDES obtain identical performance, since they implement the same detector (\ref{eq:test_CDESp}). 

In Fig. \ref{fig:Prob_sensing_vsC}, the  sensing accuracy  (\ref{eq:sa}) of all schemes is plotted versus the available fronthaul capacity $\bar{C}$ for $K=3$, $B_{th}=2$ and $P_T=23$ dB. Fig. \ref{fig:Prob_sensing_vsC} shows that, in the low fronthaul capacity regime, e.g., for $\bar{C} \le 3$, edge-based sensing is preferable, with EDCS outperforming CDCS unless the fronthaul capacity is large enough, e.g., $\bar{C} \ge 30$.

In Fig. \ref{fig:Prob_sensing_vsK}, we illustrate the sensing accuracy versus the number $K$ of APs with $\bar{C}=4$, $B_{th}=6$, and $P_T=23$ dB. Thanks to the spatial diversity afforded by multiple APs, the sensing accuracy of all functional splits improves with the number $K$ of APs. However, the improvements in sensing accuracy for edge-based sensing saturates with $K$, while that of cloud-based sensing increases at a faster rate. This is because the performance gain afforded by the majority rule is relatively smaller than that obtained by cloud-based sensing as the number of APs increases.

\subsection{Cloud vs. Edge Decoding}\label{sec:CE_decoding}
\begin{figure}[t]
\centering
\includegraphics[width=12cm]{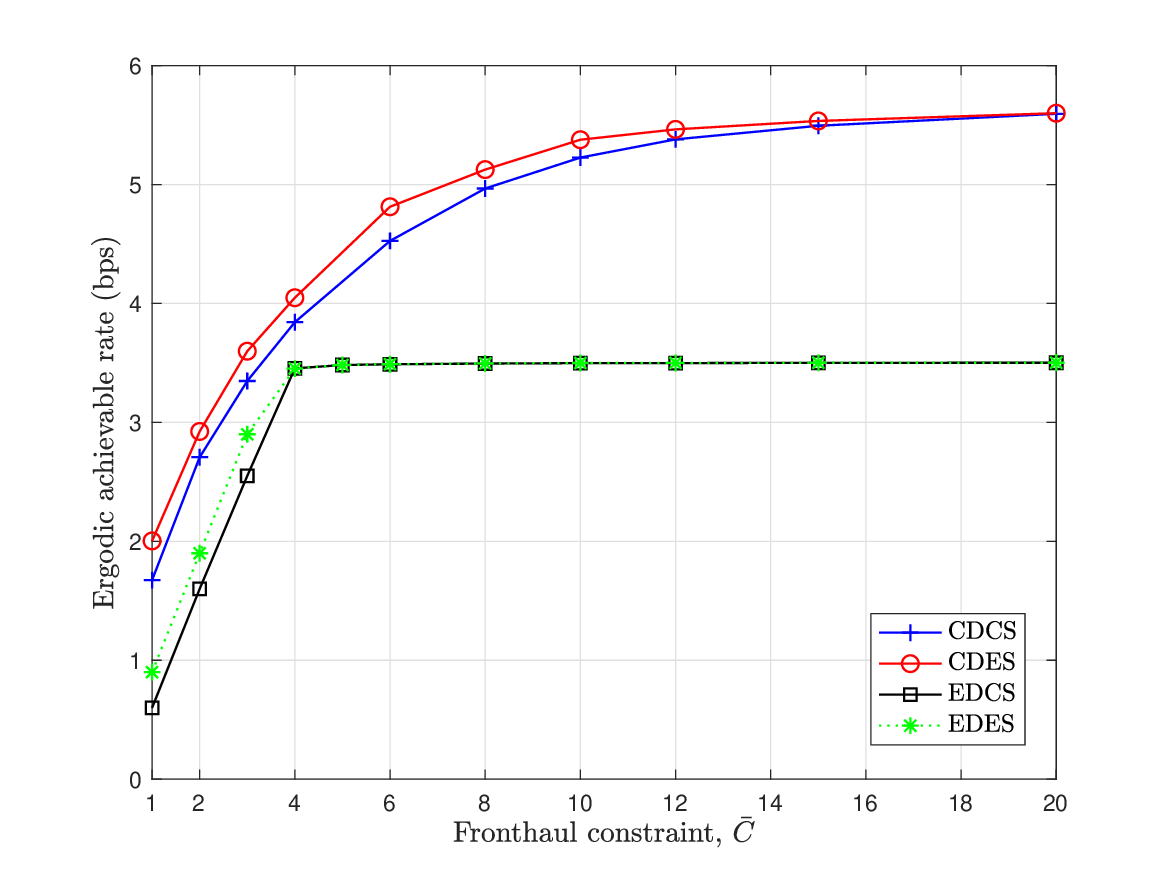}
\caption{Ergodic achievable rate versus fronthaul capacity constraint $\bar{C}$ ($K=3$, $B_{th}=2$, and $P_T=23$ dB)}
\label{fig:rate_vsC}
\end{figure}

We now turn to the performance in terms of communication rates. Fig. \ref{fig:rate_vsC} shows the ergodic achievable rate in the same condition of Fig. \ref{fig:Prob_sensing_vsC}. Cloud-based decoding is generally advantageous - particularly when coupled with edge-based sensing, which requires minimum fronthaul for sensing. In the case of sufficiently large fronthaul capacity, the achievable rate of cloud-based sensing approaches to that of edge-based sensing with the same decoding scheme. 

\begin{figure}[h!]
\centering
\includegraphics[width=12cm]{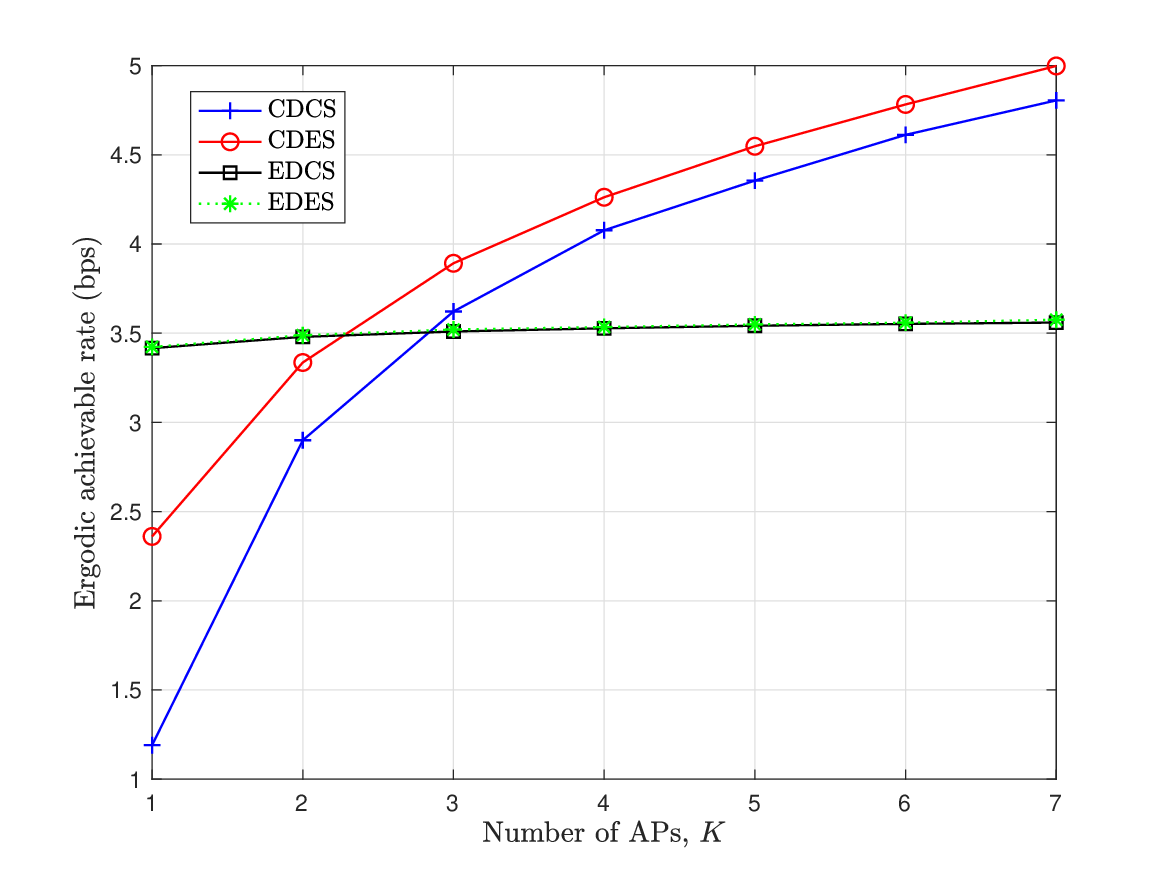}
\caption{Ergodic achievable rate versus the number $K$ of APs  ($\bar C=4$, $B_{th}=2$, and $P_T=23$ dB)}
\label{fig:rate_vsK}
\end{figure} 

Fig. \ref{fig:rate_vsK} shows the achievable rate of versus the number $K$ of APs under the same condition of \ref{fig:Prob_sensing_vsK}. Thanks to spatial diversity, the achievable rates of both edge-based and cloud-based decoding increase although the growth is faster for cloud-based decoding. Conversely, edge-based decoding via EDCS or EDES is preferred for very small values of $K$.

\subsection{Cloud vs. Edge Sensing and Communication}\label{sec:CE} 
\begin{figure}[h!]
\centering
\includegraphics[width=12cm]{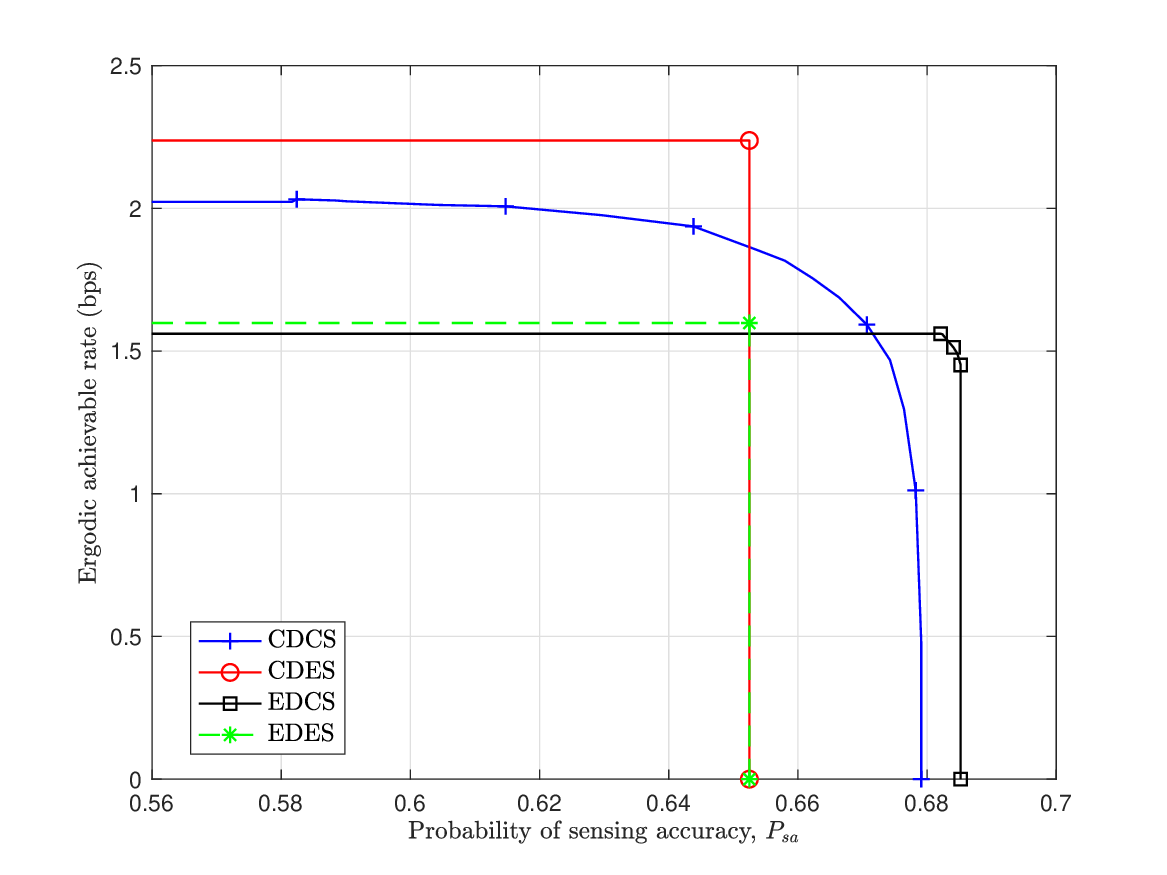}
\caption{Ergodic achievable rate versus sensing accuracy $P_{sa}$($\bar{C}=2$, $K=3$, and $P_T=15$ dB)}
\label{fig:rate_vsProb_sensing}
\end{figure} 

Finally, we analyze the trade-off between sensing and communication by plotting the achievable ergodic rate versus the sensing accuracy for $\bar{C}=2$, $K=3$, and $P_T=15$ dB in Fig. \ref{fig:rate_vsProb_sensing}. The curves in Fig. \ref{fig:rate_vsProb_sensing} are obtained by following the design problem (\ref{eq:CDCS1}), (\ref{eq:CDES1}), (\ref{eq:EDCS1}) and (\ref{eq:EDES1}) by varying the sensing constraint $B_{\text{th}}$. Edge-based sensing schemes are limited in the extent to which they can increase the sensing accuracy. However, when the sensing requirements are not too significant, edge-based sensing supports larger communication rates. In particular, cloud-based sensing schemes can achieve a larger sensing accuracy, but at the cost of decreasing achievable rates. The higher sensing accuracy is offered by EDCS, which entails a lower communication rate as compared to CDCS. Conversely, the largest communication rate is guaranteed by CDES, which devotes most of the fronthaul capacity to communication signals.

\section{Conclusions}\label{sec:con}
In this paper, we have studied the optimal fronthaul design for joint communication and radar sensing in perceptive mobile networks based on cell-free MIMO architectures, where the distributed access points and central processing unit are connected via constrained fronthaul links. Four types of functional splits are studied: $(i)$ cloud-based decoding and sensing (CDCS), $(ii)$ cloud-based decoding and edge-based sensing (CDES), $(iii)$ edge-based decoding and cloud-based sensing (EDCS) and $(iv)$ edge-based decoding and sensing (EDES). For all four types of functional splits, we have provided a unified design to optimize the fronthaul quantization of received signals during training and data phases. 

Via simulations, the trade-off between data communication and radar sensing was evaluated in the different PMN scenarios. In terms of sensing accuracy, cloud sensing supports the best radar performance. However, when the sensing requirements are limited, CDES offers a suitable solution that maximizes the communication rate. For larger sensing accuracy, CDCS or EDCS is necessary, with the latter attaining larger communication rates. It can be concluded that the choice of the functional split depends on the designed trade-off between communication and radar sensing. Interesting direction for future work include studying the estimation of radar parameters such as location, delay spread, and Doppler, as well as the consideration of multiple sensing targets and multiple communication users. 

\newpage
\begin{appendices}
\section{Fronthaul optimization for CDCS}\label{app: alg}
In this appendix, we have provided the heuristic algorithm of CDCS to obtain the fronthaul optimization of problem (\ref{eq:CDCS1}) based on line search discussed in Sec. \ref{sec:CDCS}. 

\begin{algorithm}
\caption{Fronthaul optimization for CDCS}\label{alg:CDCS}
\begin{algorithmic}
\State {\bf{Input}}: Error parameter $\epsilon \ge 0$
\State Initialize $C_{p,k}^{\text{up}}(\sigma_{p,k}^2)=0$ and $C_{d,k}^{\text{up}}(\sigma_{d,k}^2) = \bar{C}_k - C_{p,k}^{\text{up}}(\sigma_{p,k}^2)$, for all $k$ and $R^*=R_{\text{temp}}=0$. 
\State Calculate the corresponding $\sigma_{p,k}^2$ and $\sigma_{d,k}^2$ by using (\ref{eq:Ck_CDCS_front}), for all $k$.
\While{$C_{p,k}^{\text{up}}(\sigma_{p,k}^2) \le \bar{C}_k$}
\If{$\mathcal{B}(\pmb{\sigma}_p^2) \ge B_{\text{th}}$}
    \State Calculate the sum-rate in (\ref{eq:obj_CDCS1}), and set $R_{\text{temp}}$ to (\ref{eq:obj_CDCS1}).
    \If{$R_{\text{temp}} > R^*$}
    		\State $C_{p,k}^{\text{up}*}(\sigma_{p,k}^2) \leftarrow C_{p,k}^{\text{up}}(\sigma_{p,k}^2)$
    		\State $C_{d,k}^{\text{up}*}(\sigma_{d,k}^2) \leftarrow C_{d,k}^{\text{up}}(\sigma_{d,k}^2)$
    		\State $R^* \leftarrow R_{temp}$
    		\State Update the corresponding $\{\sigma_{p,k}^2\}_{\forall k}$ and $\{\sigma_{d,k}^2\}_{\forall k}$ to $\pmb{\sigma}_{p}^*$ and $\pmb{\sigma}_{p}^*$.  
    \EndIf
\EndIf
\State $C_{p,k}^{\text{up}}(\sigma_{p,k}^2) \leftarrow C_{p,k}^{\text{up}}(\sigma_{p,k}^2) + \epsilon$
\State $C_{d,k}^{\text{up}}(\sigma_{d,k}^2) \leftarrow C_{d,k}^{\text{up}}(\sigma_{d,k}^2) - \epsilon$ 
\State Update the corresponding $\sigma_{p,k}^2$ and $\sigma_{d,k}^2$.  
\EndWhile
\State {\bf{Output}}: $\pmb{\sigma}_p^*$ and $\pmb{\sigma}_d^*$
\end{algorithmic}
\end{algorithm}
\end{appendices}

\bibliographystyle{IEEEtran}
\bibliography{JSAref}

\end{document}